\documentclass[12pt]{iopart}
\usepackage{graphicx}
\usepackage{iopams}
\usepackage[section]{placeins}   
\usepackage{hyperref}  
\usepackage{subfigure}
\hypersetup{colorlinks,urlcolor=blue,linkcolor=blue,citecolor=blue,filecolor=blue,urlcolor=blue}
\usepackage{ifthen}
\usepackage{cite}
\usepackage{booktabs}
\usepackage{multirow}

\begin{document}

\markboth{Adriano Silva, Sergio Floquet and Ricardo Lima}{Newcomb-Benford Law in Neuromuscular Transmission: Validation at Hyperkalemic Conditions}

\title[Newcomb-Benford Law in Neuromuscular Transmission]{Newcomb-Benford Law in Neuromuscular Transmission: Validation at Hyperkalemic Conditions}

\author{Adriano Silva$^{1}$, Sergio Floquet$^{2}$ and Ricardo Lima$^{3}$}
\address{$^{1}$ Universidade Federal do Sul da Bahia, Itabuna, BA 45613-204, Brazil}
\address{$^{2}$ Colegiado de Engenharia Civil, Universidade Federal do Vale do S\~ao Francisco, Juazeiro, BA 48902-300, Brazil}
\address{$^{3}$ Departamento de Fisiologia e Farmacologia, Faculdade de Medicina, Universidade Federal do Cear\'a, Fortaleza, CE 60430-270, Brazil}
\ead{adjesbrl@ufsb.edu.br}

\date{\today}

\begin{abstract}
Recently, we demonstrated the validity of anomalous numbers law, known as Newcomb-Benford´s law, at the mammalian neuromuscular transmission, considering different extracellular calcium concentrations \cite{silva2020}. The present work continues to examine how alterations in extracellular solution modulate the first digit law in the context of the spontaneous release of acetylcholine from the neuromuscular junction. We investigated if the intervals of miniature potentials collected at the neuromuscular junction obey the law in a hyperkalemic environment. The analysis showed that the interval between the miniature potentials at high potassium concentrations follows Newcomb-Benford´s law. Also, our data allowed us to uncover a conformity fluctuation as a function of the number of intervals of the miniature potentials. Finally, we discuss the biophysical implications of the present findings.

\vspace{0.5cm}
\noindent
{\it Keywords\/}:{electrophysiology; neuromuscular junction; Newcomb-Benford´s law; time series; potassium}
\end{abstract}

\section{Introduction}

The neuromuscular junction ($NMJ$) is a specialized region that establishes communication between nerve and muscle. The language of this communication is chemical, in which acetylcholine molecules, packed inside organelles called vesicles, are released after their fusion into the synaptic cleft. Next, a diffusion occurs within the synaptic cleft where acetylcholine binds to cholinergic receptors in the motor end-plate, promoting a muscular response \cite{kloot1991}. Thanks to the extensive work of Katz and collaborators, his group systematically carried out rigorous characterization work from the 1950s onwards. In conjunction with these investigations, the discovery of miniature end-plate potentials ($MEPPs$) by Katz and Fatt represented a new perspective for understanding the biophysical nature of neurotransmission \cite{katz1971}. Would there be a way to quantify the spontaneous release of acetylcholine? According to the vesicular hypothesis proposed later by Katz, the release took place discreetly in the terminal, where there would be a direct correspondence between the fusion of a single vesicle and the generation of a $MEPP$ \cite{bennett2000}. Therefore, based on these studies, Katz and Del Castillo offered a statistical pillar consistent with the physiological substrate. According to this proposal, the release occurred within a random regime, statistically governed by a Poisson regime.

The technological improvement allowed the development of electrophysiological instrumentation, raising the quality of the records. In conjunction with these empirical advances, sophisticated statistical models were also proposed, enabling the test of the validity of the Poissonian premises. Within this perspective, several studies have emerged showing a divergent scheme concerning the assumptions based on the randomness of the neurotransmission phenomenon. Especially from the 1970s onwards, several authors showed that neurotransmission could also obey other statistical models. For instance, Robinson, examining the $NMJ$ of amphibians, showed that $MEPPs$ was best described when analyzed with the gamma function \cite{robinson1976}. Washio, studying cockroach neurojunctions, also observed deviations from the Poisson statistics, while van der Kloot, still examining the amphibian $NMJ$, also reinforced such divergences \cite{washio1980,cohen1973}. More recently, Lowen \textit{et al.}, and independently Takeda \textit{et al.}, suggested a fractality acting on cholinergic release \cite{lowen1998,takeda1999}. In addition, morphological studies have shown that the physical interaction between the vesicles can generate both inhibition and exacerbation of vesicular release \cite{harlow2001,bennett1990}. This finding encourages the idea of mechanisms based on long-range interactions acting on the process responsible for the release. In this context, emerged the possibility for such mechanisms to be linked to a nonextensivity regime as documented by Silva \textit{et al.} \cite{silva2020}. These authors suggested the q-Gaussian distribution as an adequate function to adjust $MEPPs$ amplitudes. Therefore, these results indirectly provide support for the morphological investigations mentioned above.

Numerical patterns are identified in many phenomena of nature. Like so many other discoveries in the history of science, the
Newcomb-Beford´s Law 
was obtained in a completely unexpected manner. It was the result of an accurate examination of logarithm tables. Using them, Simon Newcomb discovered a curious numerical pattern in $1881$ \cite{newcomb1881}. Newcomb was a respectable researcher known for his remarkable skill in dealing with large amounts of data. His accurate examination allowed him, when handling a catalog with logarithmic tables worn out by routine consultation, to notice that the initial pages of the catalog, containing the number $1$ as the first digit, were more worn out than the pages containing the digit $2$, and so on. From these observations, he proposed a simple mathematical equation to explain the curious phenomenon, also known as the law of anomalous numbers or the law of the first digit, represented by an elegant logarithmic function.

After decades, Frank Benford independently rediscovered Newcomb findings, analyzing a large amount of data extracted from various sources, such as physical constants, molecular weights, and the height of the American population. By studying them, Benford reached the same conclusions previously pointed out by Newcomb \cite{benford1938}. Since then, the law of first digits has been known as Benford's Law, therefore, to do justice to Newcomb, we will name it by Newcomb-Beford´s Law ($NBL$). The $NBL$ is classified among the several power or scaling laws in many physical systems. In recent years, it has emerged as a valuable tool for identifying patterns embedded in data from different data sources \cite{kossovsky2014}. Nevertheless, $NBL$ remained a curious mathematical observation until a rigorous treatment explained why it works so well in different phenomena. Despite the functional simplicity of the law, researchers still need to understand why $NBL$ works so well \cite{berger2015}. In the middle of the 1990s, Hill \cite{hill1995a,hill1995b,hill1995c} offered a formal understanding of two remarkable characteristics of the law: scale and basis invariance.

Various experimental data has been accumulated in many fields of knowledge, attesting compliance with $NBL$ \cite{burgos2021,morag2019,crocetti2016,yan2018,toledo2015}. In particular, spatial invariance can have profound morphological implications in physiology, and it is well documented in the heart, lung, and brain \cite{bassingthwaighte1994}. In this framework, it is plausible to hypothesize that if a given data collection obeys the $NBL$ then it should exhibit base invariance behavior. Therefore, it is unsurprising that $NBL$ has been confirmed in several biological systems. Studies carried out on electrocardiogram and electroencephalogram recordings have revealed to follow the $NBL$ in the physiological level \cite{seenivasan2016,tirunagari2017}. Moreover, \textit{in vitro} studies, using the $NMJ$ of mouse diaphragms, Silva \textit{et al.} performed a detailed electrophysiological investigation into the validity of the law at different extracellular calcium levels \cite{silva2020}. According to these authors, the intervals between $MEPPs$ obey the $NBL$, no matter the calcium concentration, indicating robust conformity of their data with the law. Motivated by this work, it is suitable to delve into investigations varying the extracellular content of other ionic species.

The potassium ions ($K^{+}$) is a univalent cation commonly found in corporal fluids, resting within $3.5-5$ mM, being crucial for several physiological functions \cite{lindinger2021}. For example, changes in $[K^{+}]_{o}$ gradient represent a potential risk for cardiac functions, and it is also known for establishing the $K^{+}$-equilibrium potential, which is vital for several cell functions. Beyond that, the membrane potential depolarization due to high $[K^{+}]_{o}$, implies a dramatic increase in $MEPP$ frequency. The $K^{+}$ traffic is mediated by several channels at the $NMJ$ membrane terminal \cite{maljevic2013}. Thus, beyond the health issues, modifications in $K^{+}$ content in the extracellular milieu represent fertile soil for mathematical modeling of the ionic impact on the nervous electrical activity. In this framework, the $NMJ$ emerges as a classical but still essential preparation to identify numerical patterns in a biological scope. Deviations from the $[K^{+}]_{o}$ could eventually represent an opportunity for uncovering specific numerical patterns associated with determined pathological regimes playing a role in neurotransmission.

In summary, the present work is based on the manipulation of high extracellular potassium ($[K^{+}]_{o}$) because it approximately mimics a physiological stimulation \cite{grohovaz1989}. Also, impacts exerted by the manipulation of extracellular and intracellular [$K^{+}$] over the membrane potential in muscle preparations are well characterized \cite{adrian1956}. Next, $[K^{+}]_{o}$ triggers a strong membrane depolarization followed by a dramatic acceleration of the $MEPP$ rate \cite{glavinovic1988}. Third, several studies have correlated morphological cellular modifications evoked by the accumulation of $[K^{+}]_{o}$ \cite{sykova2008}. From this justification, the present work aims to expand our previous study, evaluating whether the intervals between $MEPPs$ still obey the law in conditions of hyperkalemia. It is well accepted that the increase in extracellular potassium concentration increases the frequency of $MEPPs$. Therefore, this exacerbated electrophysiological activity allows rigorous verification of the law validity for a large amount of data, and the conformity level may be studied in more detail.

\section{Mathematical formulation of NBL}

Hill introduced the probabilities for occurrences, inferred by the general equation expressed as follows:

\begin{equation}
 P( D_{1}=d_{1}, \cdots , D_{k} = d_{k} )  =  \log \left[ 1 + \frac{1}{ \displaystyle \sum_{i=1}^{k} d_{i} \times 10^{k-1} }  \right].  \label{eq-hill}
\end{equation}

The above expression can be particularized to analyze only the frequency of the first digits. In this case, the equation is then written as follows:

\begin{equation}
 P( D_{1}  = d_{1} )  =  \log \left(1 + \frac{1}{d_{1}}  \right), \qquad  d_{1} \in \left\lbrace 1,2, \cdots ,9  \right\rbrace. \label{benford}
\end{equation}

It is worth to highlight that second digit analysis is often performed in the $NBL$ applications. For example, Diekmann documented that articles published in the American Journal of Sociology are well described by taking the second digit \cite{diekmann2007}. Thus, probabilities for the appearance of a second digit are given by the expression:

\begin{equation}
 P(  D_{2} = d_{2} )  =  \sum_{d_{1}=1}^{9} \log \left( 1+  \frac{1}{d_{1}d_{2}}  \right), \qquad d_{2} \in \left\lbrace 0,1, \cdots ,9 \right\rbrace . \label{benford-2}
\end{equation}

Nigrini claims that regardless of the usual analysis of the first or second digits, providing important information about the compliance of the analyzed data, it is vital to consider the analysis of the first two digits \cite{nigrini2012}. According to this researcher, investigating the conformity of the first two digits makes it possible to extract a more detailed scenario of how the phenomena obey the law. This case is written in the following functional form:

\begin{equation}
 P( D_{1}D_{2} = d_{1}d_{2} )  =  \log \left( 1 + \frac{1}{d_{1}d_{2}}  \right), \qquad d_{1}d_{2} \in \left\lbrace 10,11, \cdots ,99 \right\rbrace . \label{benford-2b}
\end{equation}

The expected frequencies for the first and second digit is resumed in the table~\ref{tab01}:

\begin{table}[!hb]
\caption{Frequencies for the first and second digit of $NBL$.\label{tab01}}

\vspace{0.5cm} \small
\hspace{-0.8cm}\begin{tabular}{ccccccccccc}
\toprule

Digit & 0 & 1 & 2 & 3 & 4 & 5 & 6 & 7 & 8 & 9 \\
\midrule
1st & - & 0.30103 & 0.17609 & 0.12494 & 0.09691 & 0.07918 & 0.06695 & 0.05799   & 0.05115 & 0.04576 \\
2nd & 0.11968 & 0.11389 & 0.10882 & 0.10433 & 0.10031 & 0.09668 & 0.09337 & 0.09035 & 0.08757 & 0.08500 \\

\bottomrule
\end{tabular}
\end{table}

Several authors have reported that fluctuations in the empirical first digit values can also occur, although the typically asymmetric distribution of digits. This observation implies deviations between the data and the frequency values predicted by the $NBL$. Evidence corroborating these observations comes from seismic activity and cognition experiments \cite{diaz2015,gauvrit2017}. This issue motivated several authors to propose a $NBL$ generalization. In this framework, for instance, one may highlight the theoretical description introduced by Pietronero \textit{et al.} \cite{pietronero2001}. According to this author, assuming a probability distribution $P_{\alpha}(d)$ (where d is the digit(s), $D$ represents an arbitrary digit(s) and $\alpha$ is a constant exponent related to the scale proportion), one may write:

\begin{equation}
 P_{\alpha}(d)  =  \int_{d}^{d+1} D^{-\alpha} dD , \label{eq-pn}
\end{equation}
or by the differential equation:
\begin{equation}
 \frac{dP_{\alpha}(D)}{dD}  =  D^{-\alpha}.
 \label{eq-pn-diff}
\end{equation}

Solving eq. (\ref{eq-pn-diff}) results in an $\alpha$-logarithm:
\begin{eqnarray}
 P_{\alpha}(d) & = & \frac{1}{1-\alpha}\left[ \left(d+1\right)^{(1-\alpha)} - d^{(1-\alpha)}\right] \\
   & = & d^{(1-\alpha)} \ \ln_{\alpha}\left( \frac{d+1}{d} \right).
 \label{eq-pn-sol}
\end{eqnarray}

According to eq. (\ref{eq-pn-sol}), defined as generalized $NBL$ $(gNBL)$, more frequent first digits than expected by $NBL$ implies $\alpha>1$, while $\alpha<1$ means a first digit frequency below the predicted percentage. As expected, when $\alpha=1$ $NBL$ is fully recovered. Taking $n=d_{1}$,  equation (\ref{eq-pn-sol}) rewritten as:

\begin{equation}
 P_{\alpha}( d_{1} )  =  d_{1}^{1-\alpha} \ln_{\alpha}\left( \frac{d_{1}+1}{d_{1}}  \right),  \label{gbenford}
\end{equation}

From the approach developed by Pietronero \textit{et al.} \cite{pietronero2001}, it is also possible to obtain expressions for the second digit:
\begin{equation}
  P_{\alpha}( d_{2} )  =  \sum_{d_{1}=1}^{9} (d_{1}d_{2})^{1-\alpha} \ln_{\alpha}\left( \frac{d_{1}d_{2}+1}{d_{1}d_{2}}  \right)  .  \label{gbenford-2c}
\end{equation}

Finally, the generalized Benford's Probability for the first two digits is presented as:
\begin{equation}
 P_{\alpha}(d_{1}d_{2})  =  \displaystyle  \frac{ \left[ \left( d_{1}d_{2}+1 \right)^{1-\alpha}  - \left( d_{1}d_{2} \right)^{1-\alpha}     \right]  }{1-\alpha}, \ \ \ d_{1}d_{2} \in \lbrace 10,11,12,...,99 \rbrace , 
\end{equation}
normalized for each $\alpha$ value.

\section{Electrophysiological recordings}

The hemidiaphragm is a muscle that separates the thoracic from the abdominal cavity and presents several empirical advantages. One can highlight the easy identification and dissection, which facilitates muscle extraction. Another remarkable advantage is the stereotyped spontaneous electrophysiological activity. The experimental paradigm in the present work follows the same procedure used in our previous works \cite{silva2020,silva2015}. All experimental procedures in the present work were approved by the Animal Research Committee (CETEA - UFMG, protocol 073/03) \cite{lima2010}. Wild-type adult mice were euthanized by cervical dislocation, followed by diaphragms extraction, and quickly inserted into a physiological Ringer solution containing (in mM): $NaCl$ ($137$), $NaHCO_{3}$ ($26$), $KCl$ ($5$), $NaH_{2}PO_{4}$ ($1.2$), glucose ($10$), $CaCl_{2}$ ($2.4$), and $MgCl_{2}$ (1.3). The $pH$ was adjusted to 7.4 after gassing with $95\%$ $O_{2}$ and $5\%$ $CO_{2}.$

In the experiments with high $[K^{+}]_{o}$, sodium concentration was adjusted to maintain the osmotic equilibrium. The muscles were maintained in solution at least $30$ minutes before the beginning of the electrophysiological recordings, allowing recovery from the mechanical trauma of their extraction. Next, the tissues were transferred to a recording chamber continuously irrigated with fresh fluid at $2-3$ ml/min at room temperature ($T = 24 \pm 1 ^{\circ} C $). A standard intracellular recording technique was used to monitor the frequency of spontaneous $MEPP$ by inserting a micropipette at the chosen muscle fiber. Borosilicate glass microelectrodes had resistances of $8-15$ $M\Omega$ when filled with $KCl$ solution ($3 M$). A single pipette was inserted into the fiber near the end-plate region as guided by the presence of $MEPP$ with rise times $<1$ ms. The control experiments provided $22,582$ $MEPPs$ intervals extracted from $14$ recordings, whereas $691,743$ intervals were collected from $12$ experiments for $[K^{+}]_{o} = 25 \ mM$. Thus, our experimental paradigm afforded an enormous data quantity, allowing rigorous analysis. Electrophysiology Software (John Dempster, University of Strathclyde), R Language, Origin (OriginLab, Northampton, MA), and MATLAB (The MathWorks, Inc., Natick, MA) were employed for electrophysiological acquisition and data analysis.

\begin{figure}[!ht]
\centering

\includegraphics[width=9.1 cm]{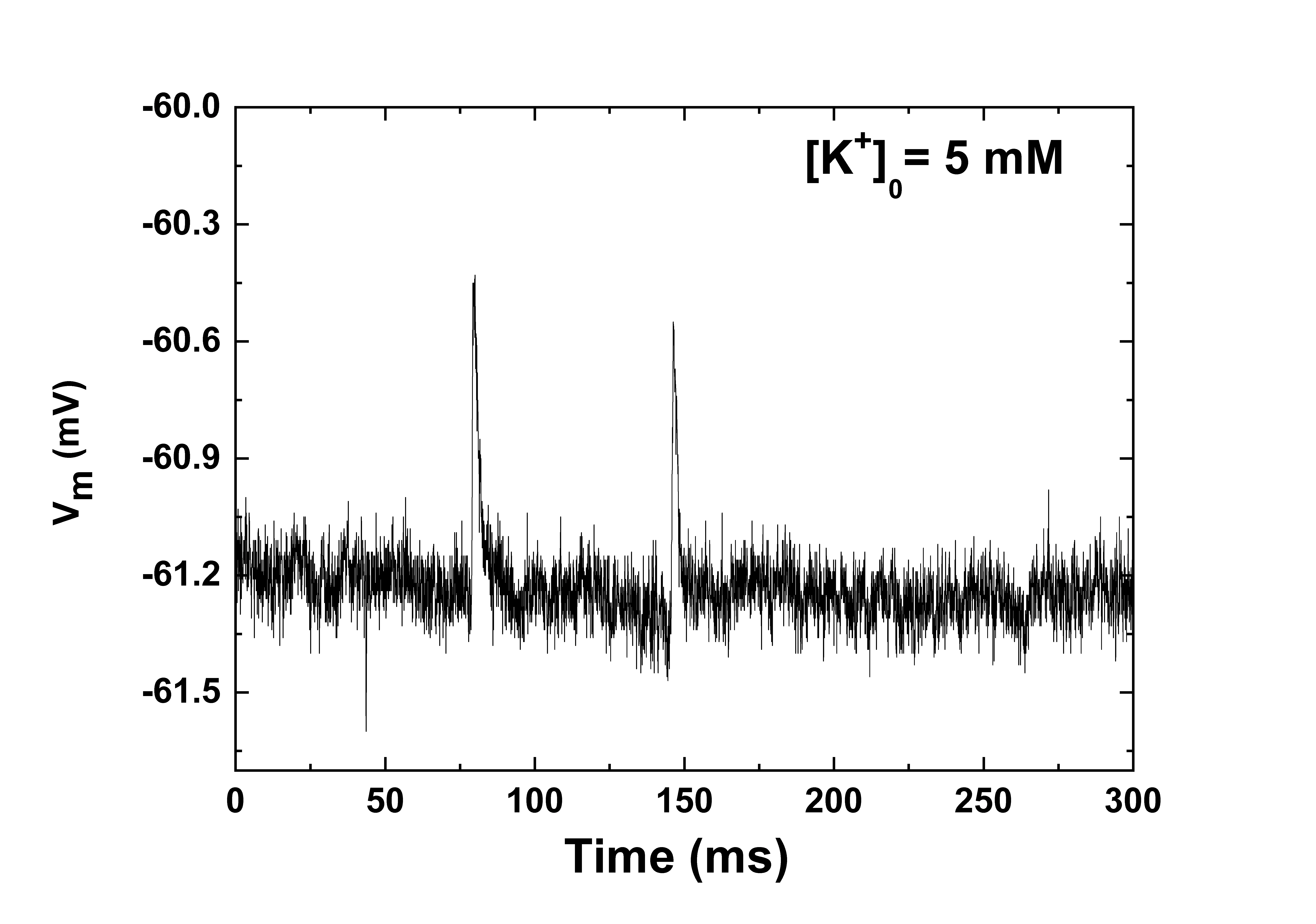}
\includegraphics[width=8.5 cm]{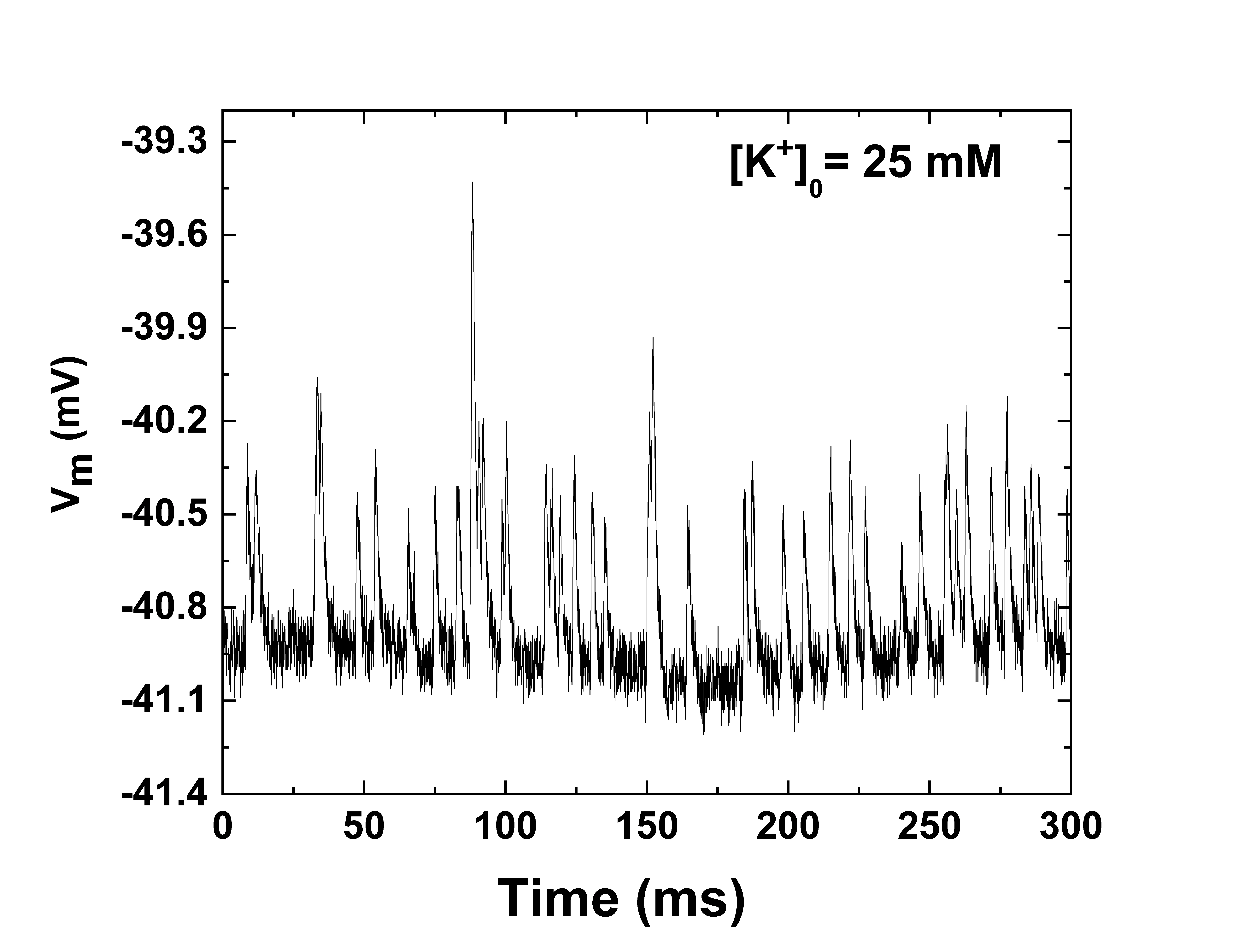}

\caption{A representative electrophysiological portions collected from two experiments carried at physiologica (left) and high $[K^{+}]_{o}$ (right).\label{fig1}}
\end{figure}

\section{Conformity analysis}

There is an intense debate about $NBL$ compliance testing. Several procedures are available, but the validity of many of these methods has been questioned. For instance, many investigators assume tests, although they manifest the "excess of power" problem, yielding in the literature accumulation of results with false claims of conformity. Adopting such tests is still more questionable when dealing with large data. In this sense, the "excess power" emerges because the tests consider the sample size in its mathematical formulation. On the other hand, while the sample size certainly confers a fundamental parameter in statistical analysis, the tests that do not consider the sample size can be interpreted as a distance of the data with those frequencies predicted by the $NBL$. Within this scheme, proposals for methods arose in which the sample size was ignored, avoiding the "excess power" problem. To address this issue, Nigrini and Kossovsky suggested the Mean Absolute Deviation ($MAD$) and the Sum of Squares Difference ($SSD$), respectively \cite {nigrini2012,kossovsky2014}. Authors suggest that despite $MAD$ the importance of calculating conformity in certain situations, $SSD$ is a superior test to $MAD$. The main reason is that the $SSD$ test does not involve absolute value, a concept directly inspired by regression theory, which uses sums of squared errors. Notwithstanding conceptual differences, $MAD$ and $SSD$ are routinely applied in different investigations. In mathematical form, $MAD$ is presented as:

\begin{equation}
  MAD  =  \frac{ \displaystyle \sum_{i=1}^{n} | AP_{i} - EP_{i}  |}{n}, \label{eq_mad}
\end{equation}
where $AP$ and $EP$ are actual and expected proportion, respectively. Additionally, SSD is calculated with the following equation:
\begin{equation}
 SSD  =  \displaystyle \sum_{i=1}^{n} \left( AP_{i} - EP_{i} \right)^{2}\times 10^{4},
\end{equation}
once again, $AP$ and $EP$ are the actual and expected proportion, respectively. Table \ref{tab2} presents the conformance range for $MAD$ and $SSD$ analysis.

\begin{table}[!ht]
\caption{Levels of conformity for first, second, and first two digits \cite{silva2020}. \label{tab2}}

\vspace{0.5cm} \small
\hspace{-1.1cm}\begin{tabular}{lllllll}
			\toprule

& \multicolumn{2}{c}{First Digit}            & \multicolumn{2}{c}{Second Digit}           & \multicolumn{2}{c}{First two Digits}         \\ \midrule 
& Range              & Conformity            & Range              & Conformity            & Range                & Conformity            \\ \midrule
 \multirow{4}{*}{MAD} & $0.000$ to $0.006$ & Close       & $0.000$ to $0.008$ & Close       & $0.0000$ to $0.0012$ & Close       \\
                      & $0.006$ to $0.012$ & Acceptable  & $0.008$ to $0.010$ & Acceptable  & $0.0012$ to $0.0018$ & Acceptable  \\
                      & $0.012$ to $0.015$ & Marginal    & $0.010$ to $0.012$ & Marginal    & $0.0018$ to $0.0022$ & Marginal    \\
                      & above $0.015$      & Nonconformity         & above $0.012$      & Nonconformity         & above $0.0022$       & Nonconformity         \\ \midrule
 \multirow{4}{*}{SSD} & $0$ to $2$         & Close       & $0$ to $2$         & Close       & $0$ to $2$           & Close       \\
                      & $2$ to $25$        & Acceptable  & $2$ to $10$        & Acceptable  & $2$ to $10$          & Acceptable  \\
                      & $25$ to $100$      & Marginal    & $10$ to $50$       & Marginal    & $10$ to $50$         & Marginal    \\
                      & above $100$        & Nonconformity         & above $50$         & Nonconformity         & above $50$           & Nonconformity         \\

			\bottomrule
		\end{tabular}
\end{table}

Recent studies showed that even the $MAD$ has inaccuracies, which will consequently reflect the actual compliance level. Investigating the foundations of this method, Lupi and Cerqueti addressed the inconsistencies in $MAD$ premises, allowing these researchers to give an alternative formulation about extracting the conformance level \cite{cerqueti2021,cerqueti2023}. These authors claim a test, still based on $MAD$, but consider the severity principle as useful to make adjustment in $MAD$ values. The excess $MAD$ test is presented as follows. Let us consider:

\begin{equation}
\displaystyle \sqrt{n} \frac{| AP_{i} - EP_{i}  |}{\sqrt{EP_{i}(1-EP_{i})}  }  \ \stackrel{d}{\longrightarrow} \ N\left(\sqrt{\frac{2}{\pi}}, 1 -\frac{2}{\pi}  \right),
\end{equation}
where $N$ is a normal distribuition. Furthermore, $MAD$ is given by:
\begin{equation}
\displaystyle \sqrt{n} MAD  \ \stackrel{d}{\longrightarrow} \ N\left(\sqrt{\frac{2}{\pi k^{2} }} \iota ' D \iota , \frac{1}{k^{2}}  \iota ' D R D \iota  \right), \label{eq_mad_nk}
\end{equation}
the simbol $'$ represent the transpose, $\iota$ is a k-vector of $1$, $D$ is a diagonal matrix formed by $D = diag(\sqrt{EP_{i}(1-EP_{i}) })$ and $R$ is the covariance matrix defined as:
\begin{equation}
R = \left( \begin{array}{cccc}
 r_{11} & r_{12} & \cdots & r_{1k} \\
 r_{21} & r_{22} & \cdots & r_{2k} \\
 \vdots & \vdots & \ddots & \vdots \\
 r_{k1} & r_{k2} & \cdots & r_{kk} \\ \end{array} \right),
\end{equation}
where
\begin{equation}
\displaystyle r_{ij} =  \frac{2}{\pi}\left( \rho_{ij} arcsin(\rho_{ij}) + \sqrt{1 - (\rho_{ij})^{2} } -1 \right),
\end{equation}
with
\begin{equation}
\displaystyle  \rho_{ij}  = \left\lbrace \begin{array}{ll}
\displaystyle -\sqrt{ \frac{EP_{i} EP_{j}}{(1-EP_{i})(1-EP_{j})}  } & for \ i\neq j  \\
1  & for \ i = j\end{array} \right. .
\end{equation}

For equation (\ref{eq_mad_nk}) $MAD$ clearly depends on $n$ and $k$, a fact reinforced by notation $MAD_{n,k}$  which allows measure discrepancy
of usual $MAD$ for his mean, denoted by Excess $MAD$
\begin{equation}
\displaystyle \delta_{n,k} = MAD_{n,k} - E(MAD_{n,k}).
\end{equation}

Therefore, $MAD $ method, given by (\ref{eq_mad}), is not independent of $n$, but rather it depends on $\displaystyle O(n^{-1/2})$ \cite{cerqueti2021,cerqueti2023}. Thus, we will not include the excess $MAD$ for $gNBL$ because it was only demonstrated that $MAD$, under the null of conformity with $NBL$, is approximately distributed as shown in (\ref{eq_mad_nk}) \cite{cerqueti2021,cerqueti2023}. In this sense, for $gNBL$ we will probably obtain another expression taking into account $\alpha$ values. Within this scope, subsequent investigations would delve in this problem in order to obtain a generalization of the $MAD$ for the $gNBL$, providing a $MAD$ as a function of $\alpha$, $k$ and $n$ ($MAD_{\alpha, k, n}$), making possible to analyze its excess $MAD$ as $\delta_{\alpha, k, n}$.

\section{Results}

\begin{figure}[!hb]
\centering
\hspace{-5.2cm} a)  \hspace{8.cm} b)

\includegraphics[width=7.7 cm]{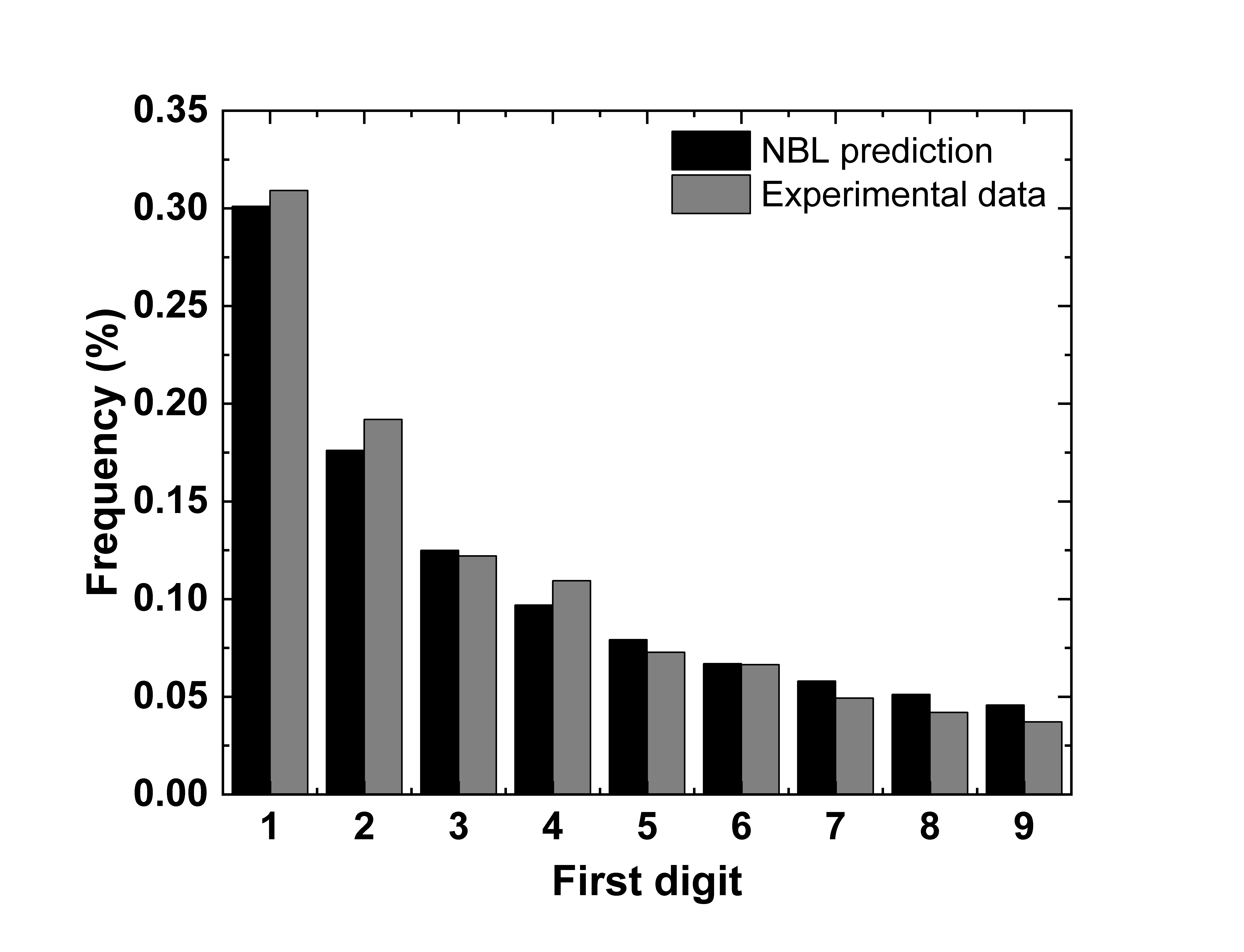}
\includegraphics[width=7.7 cm]{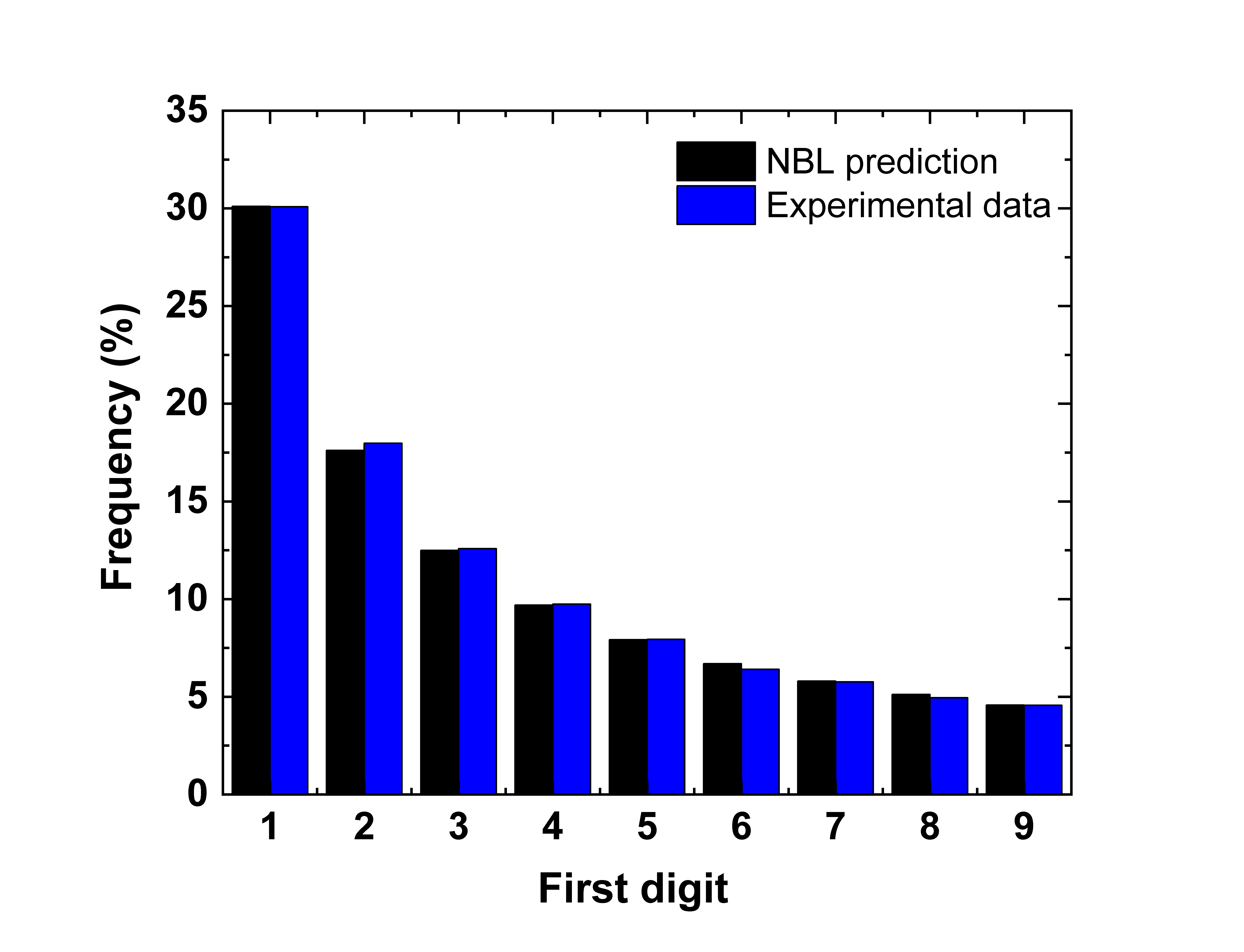}

\includegraphics[width=7.7 cm]{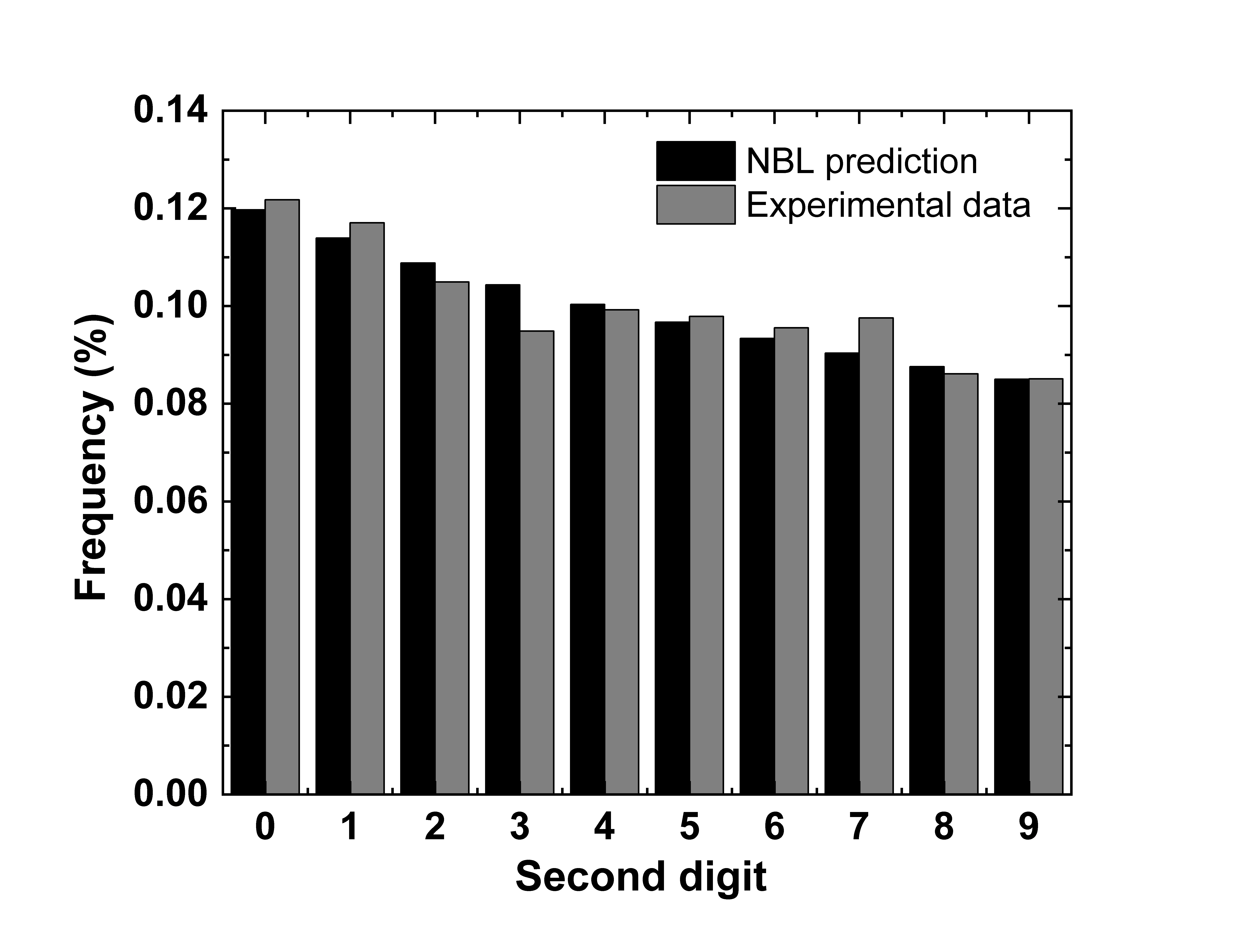}
\includegraphics[width=7.7 cm]{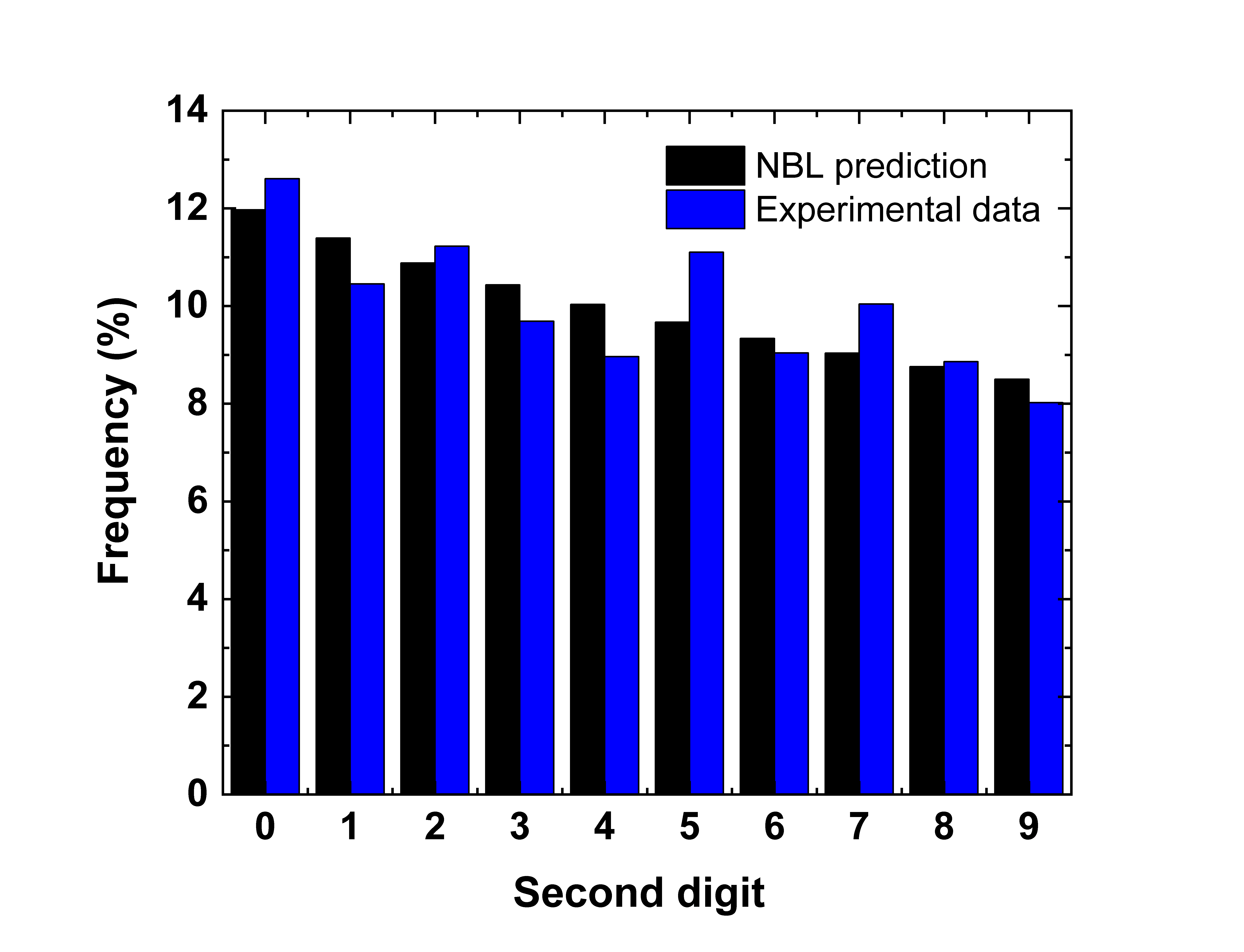}

\includegraphics[width=7.7 cm]{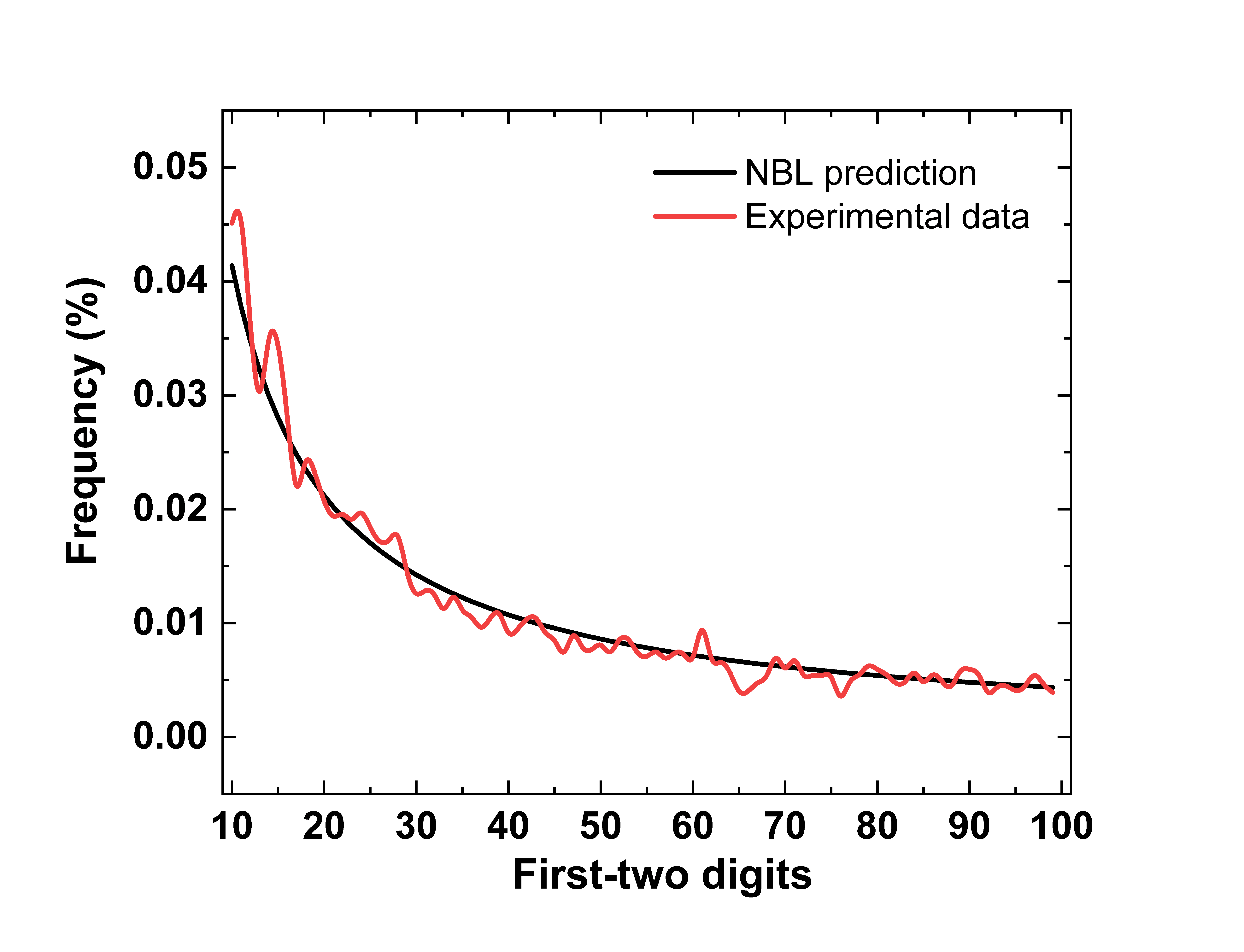}
\includegraphics[width=7.7 cm]{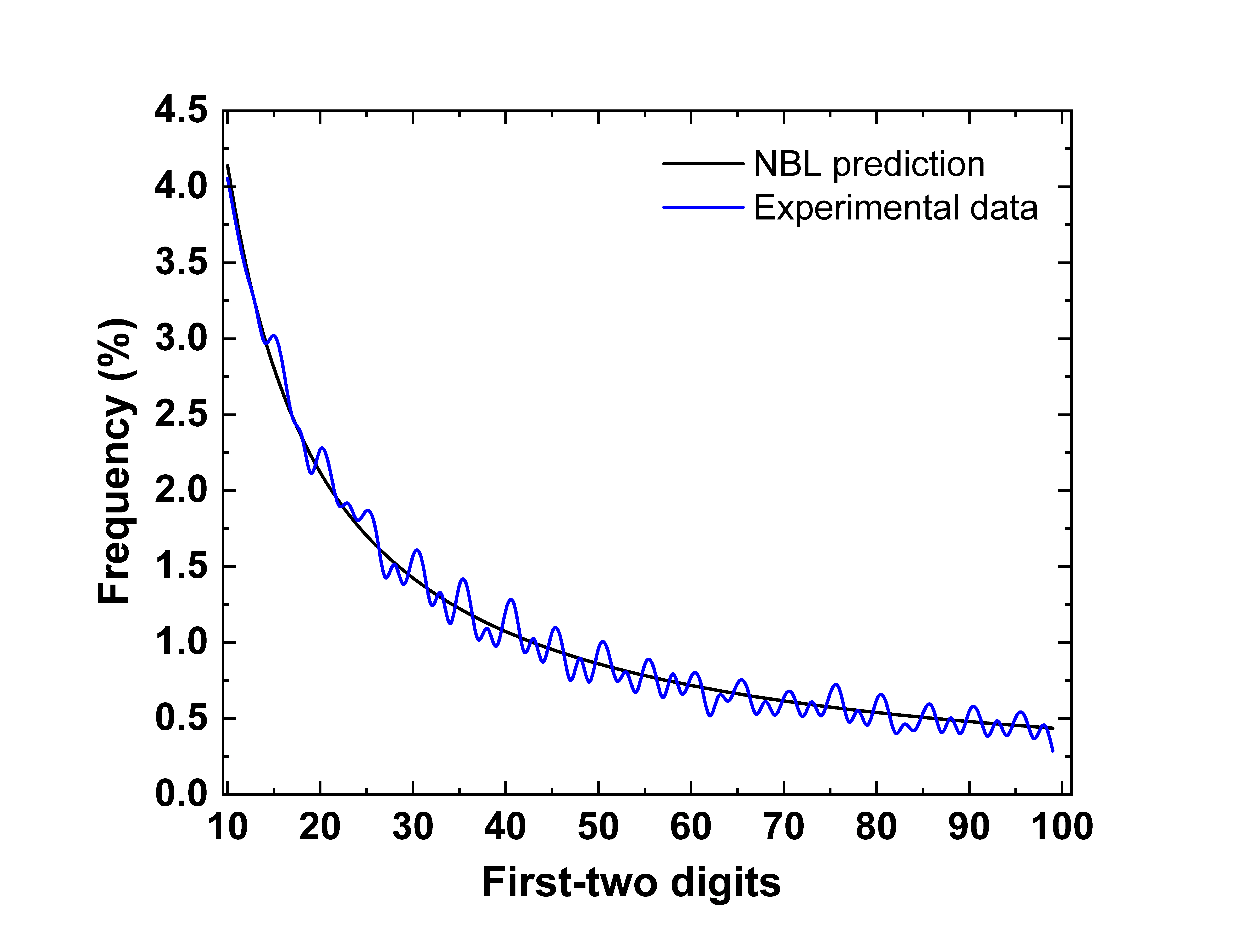}

\caption{Representative examples taken for both for $[K^{+}]_{o}$ = $5$ mM (a) and $[K^{+}]_{o}$ = $25$ mM (b), showing a satisfactory conformity between $NBL$ and experimental data. To enhance clarity the first two digits graphs are represented as lines.\label{fig2}}
\end{figure}

\begin{figure}[!hb]
\centering
\hspace{-5.2cm} a)  \hspace{8.cm} b)

\includegraphics[width=7.7 cm]{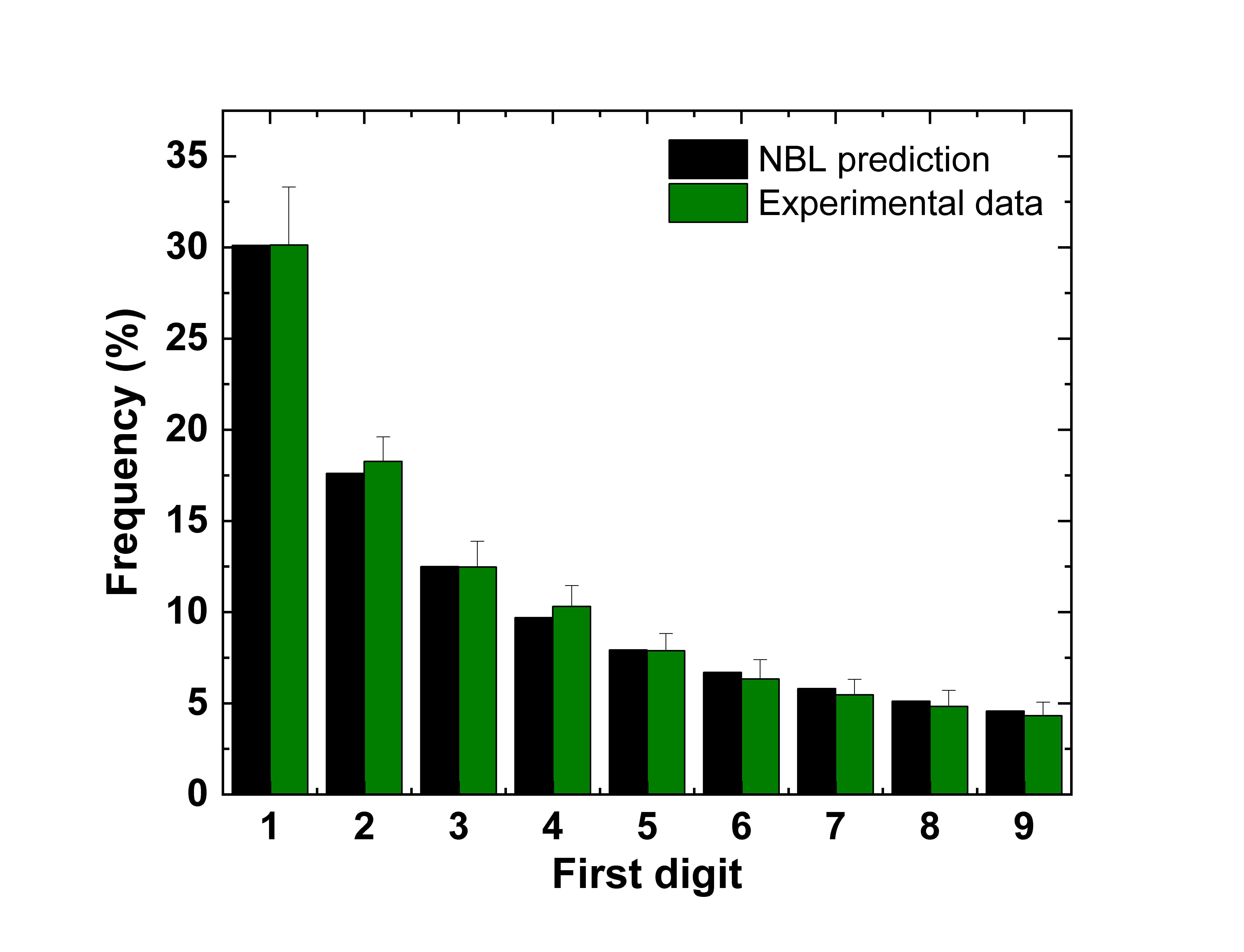}
\includegraphics[width=7.7 cm]{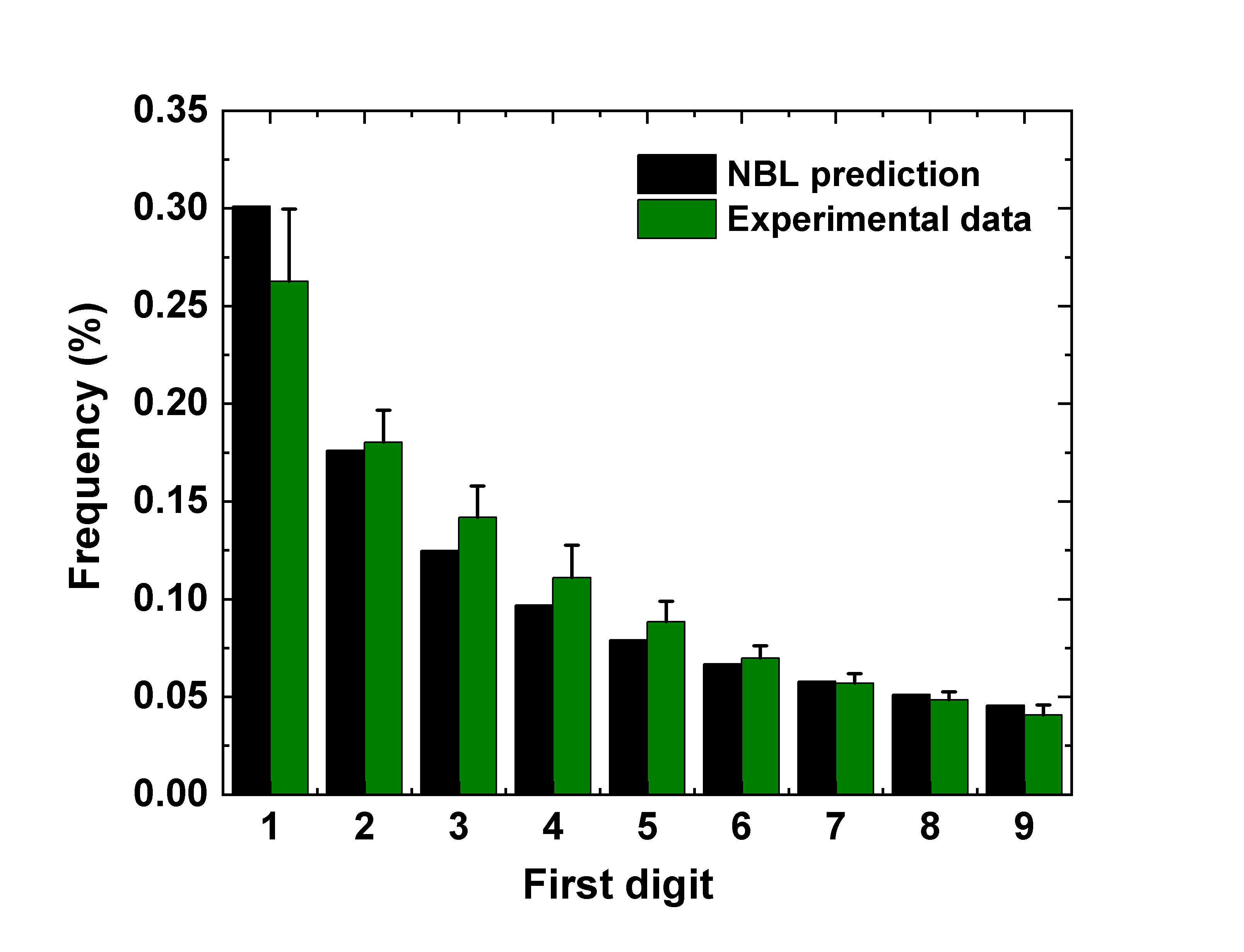}

\includegraphics[width=7.7 cm]{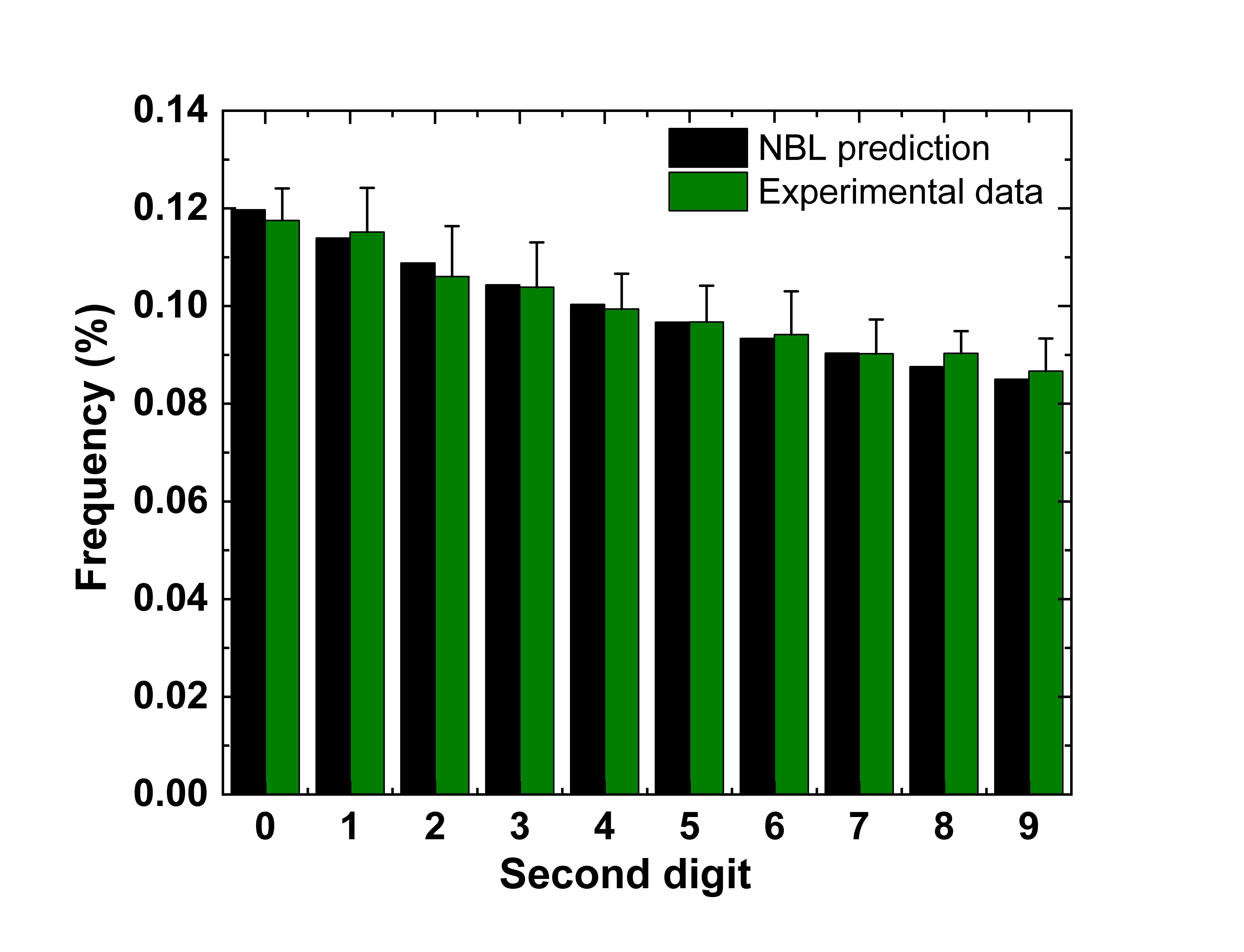}
\includegraphics[width=7.7 cm]{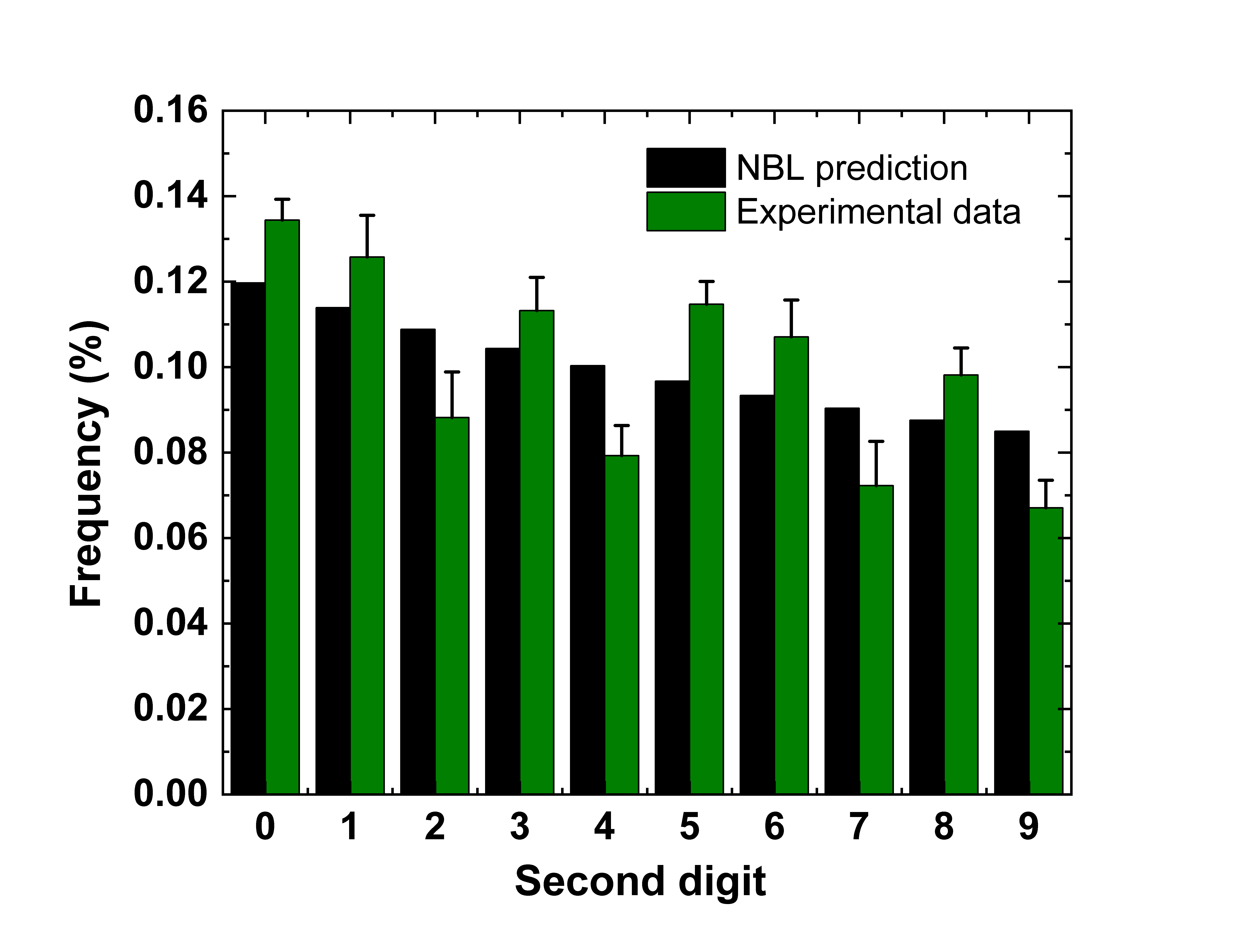}

\includegraphics[width=7.7 cm]{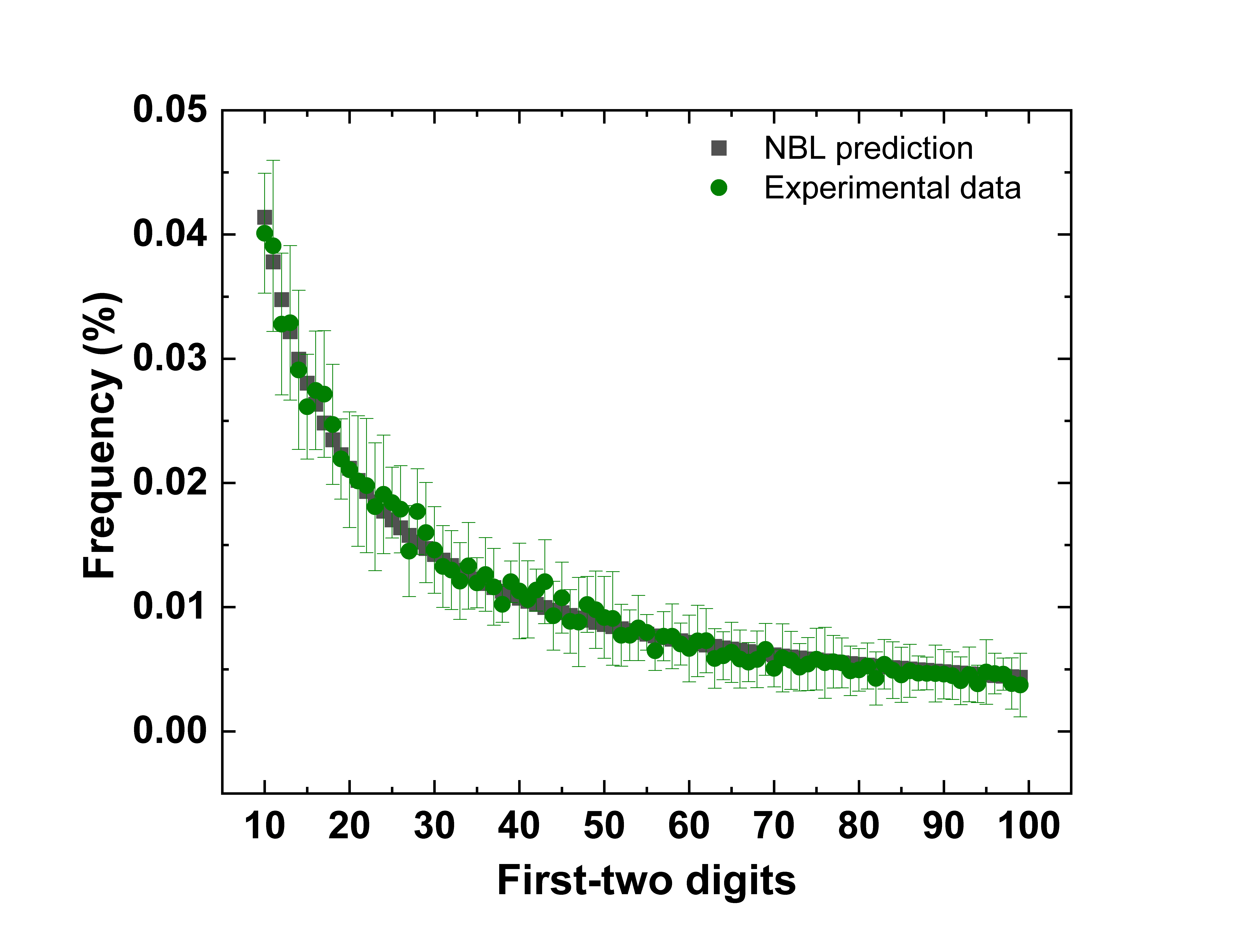}
\includegraphics[width=7.7 cm]{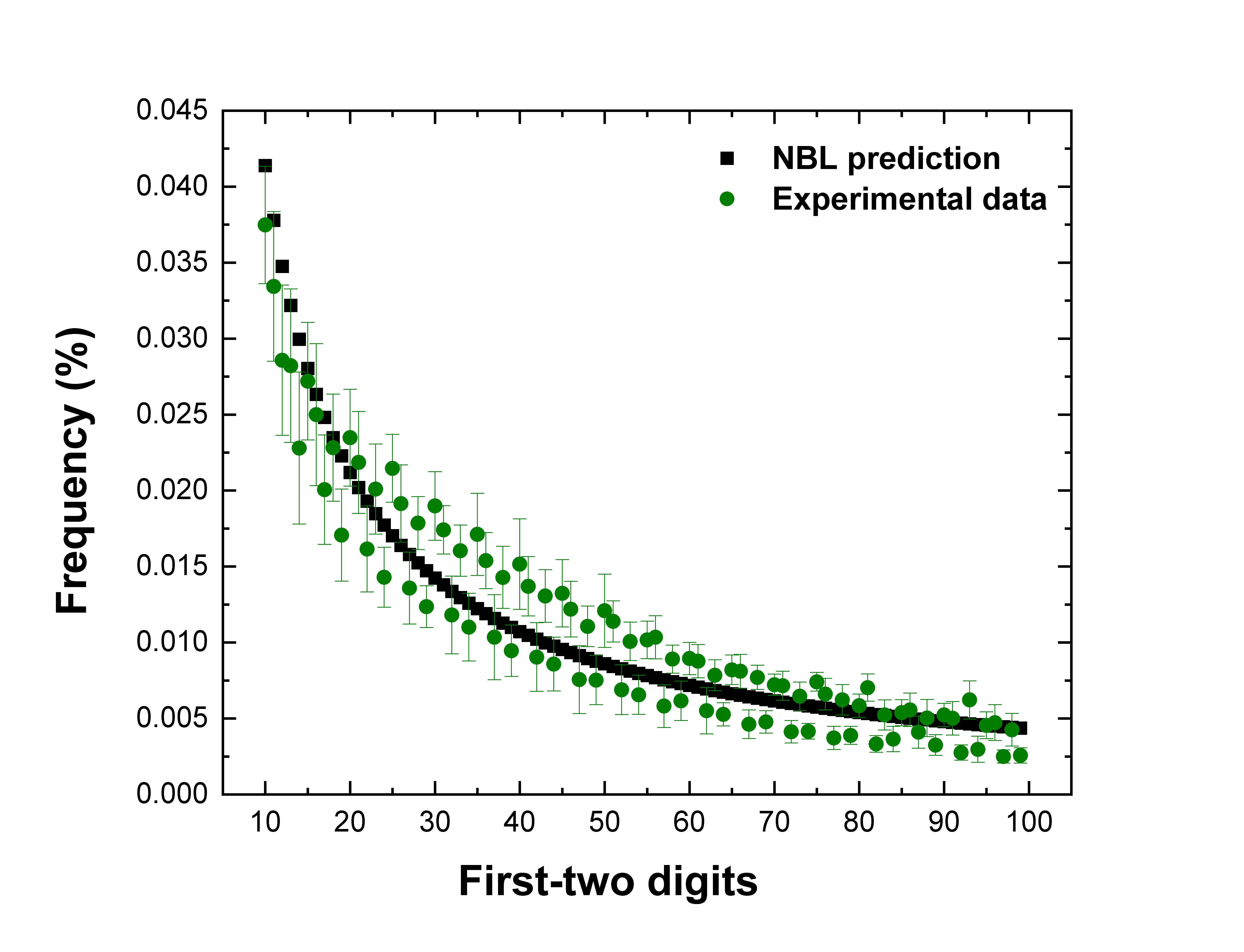}

\caption{Statistical summary for $[K^{+}]_{o}$ = $5$ mM (n = $14$) (a) and $[K^{+}]_{o}$ = $25$ mM (n = 12) (b), using $NBL$. Values are expressed as mean $\pm$ standard deviation.\label{fig3}}
\end{figure}

The conformity analysis summarized in Table \ref{tab3}, showed that at normal $[K^{+}]_{o}$ the experimental $MEPP$ intervals follow the $NBL$ satisfactorily, considering the first and second digits, as much as the first two digits. All data achieved at least marginal conformity. Moreover, it is essential to highlight that the $MAD$ excess calculation adjusted the compliance levels, especially for the first two digits, improving conformity in all cases. The excess $MAD$ possibly attenuated the "excess power", while the non-conformities obtained even with the excess $MAD$ may suggest that $NBL$ in its classic format is inadequate to describe the $MEPP$ intervals. This observation is also reinforced by examining the figure \ref{fig2}, where it can be observed representative results extracted from two electrophysiological recordings, considering both physiological level and high $[K^{+}]_{o}$. For the first digit, a visual inspection enables one to observe an excellent agreement of experimental data and predicted values. However, taking all results given in Table \ref{tab3} and \ref{tab4}, the conformity pattern revealed an exciting scenario, as different levels of compliance were observed regardless of the test used. Figure \ref{fig3} brings the statistical summary of all databases for both potassium contents, where, in general, it is possible to observe a more significant deviation of the data taken at high $[K^{+}]_{o}$. In spite of this more significant deviation, one may note, except for the second digit for high $[K^{+}]_{o}$, the characteristic asymmetric digits distribution predicted by the law.

The conformity levels obtained at high $[K^{+}]_{o}$, highlighted that although using $NBL$ gave more frequent non-conformities, the overall results suggest that the law remains most obeyed. Despite these results, we decided to delve our attention into these deviations from the $NBL$ proportions, especially for high $[K^{+}]_{o}$ recordings, where we verified whether the $NBL$ generalization might be more appropriate in the data adjustments. In this scheme, the results in Table \ref{tab5} shows for $[K^{+}]_{o} = 5 \ mM$ in $gNBL$ is assumed, it generally implies conformity improvement. This observation is readily confirmed by comparing the $SSD$ results of tables \ref{tab3} and \ref{tab5}. The conformity analysis for $[K^{+}]_{o} = 25 \ mM$, summarized in tables \ref{tab4} and \ref{tab6}, also suggested a better adherence of $gNBL$ to the data as compared to $NBL$ calculations. In fact, according to these results, the compliance level improved in several cases.

\begin{figure}[!hb]
\centering
\hspace{-5.2cm} a)  \hspace{8.cm} b)

\includegraphics[width=7.7 cm]{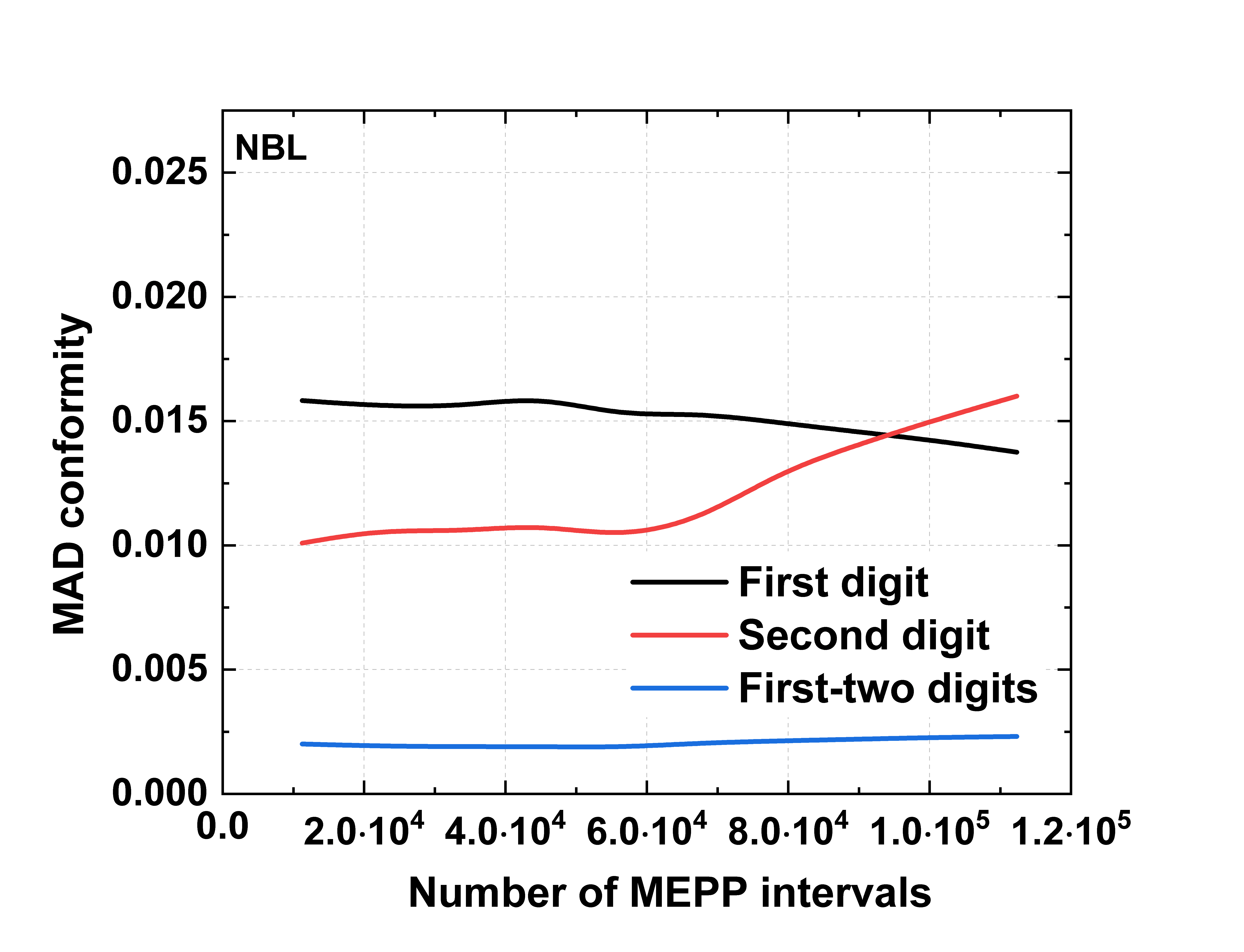}
\includegraphics[width=7.7 cm]{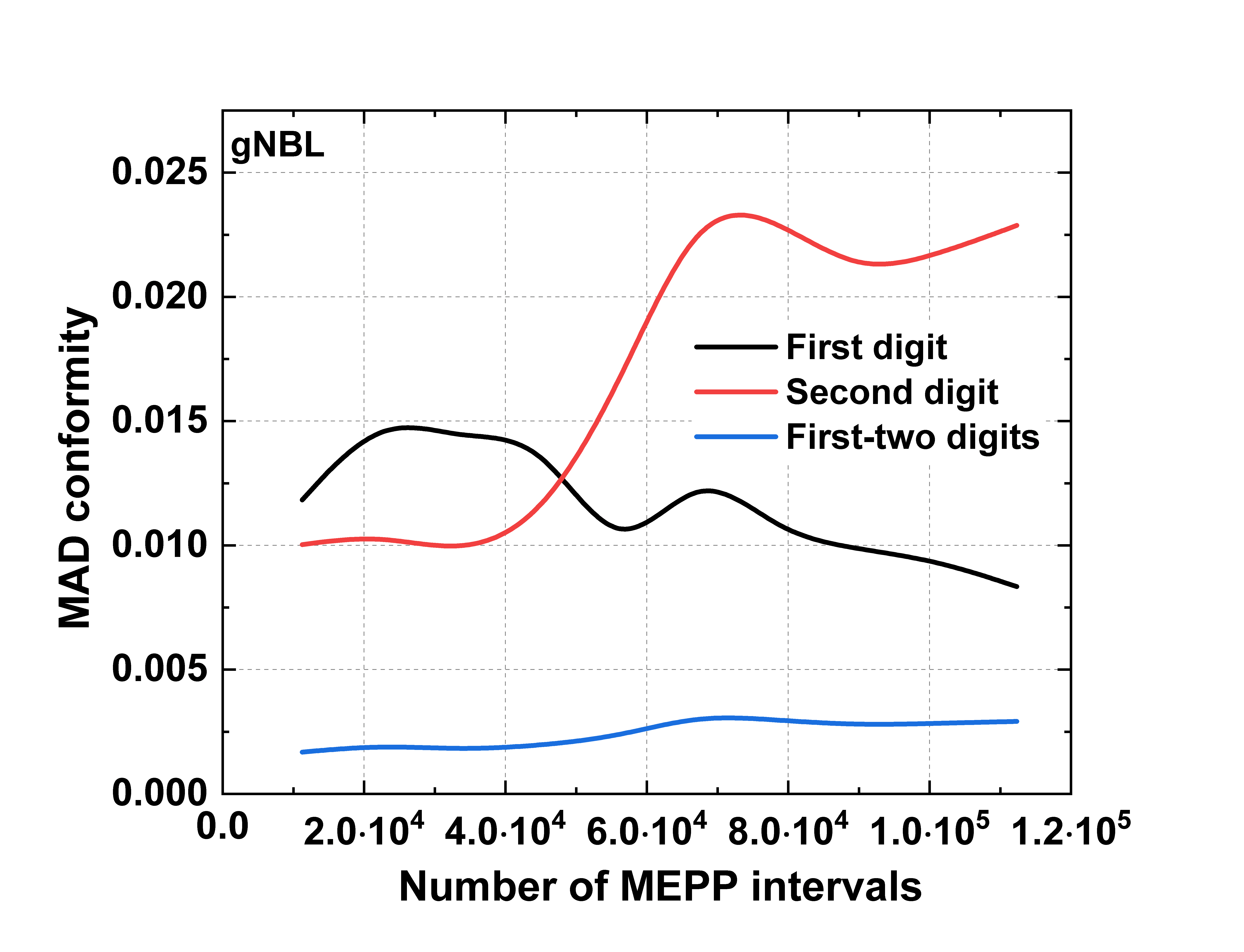}

\includegraphics[width=7.7 cm]{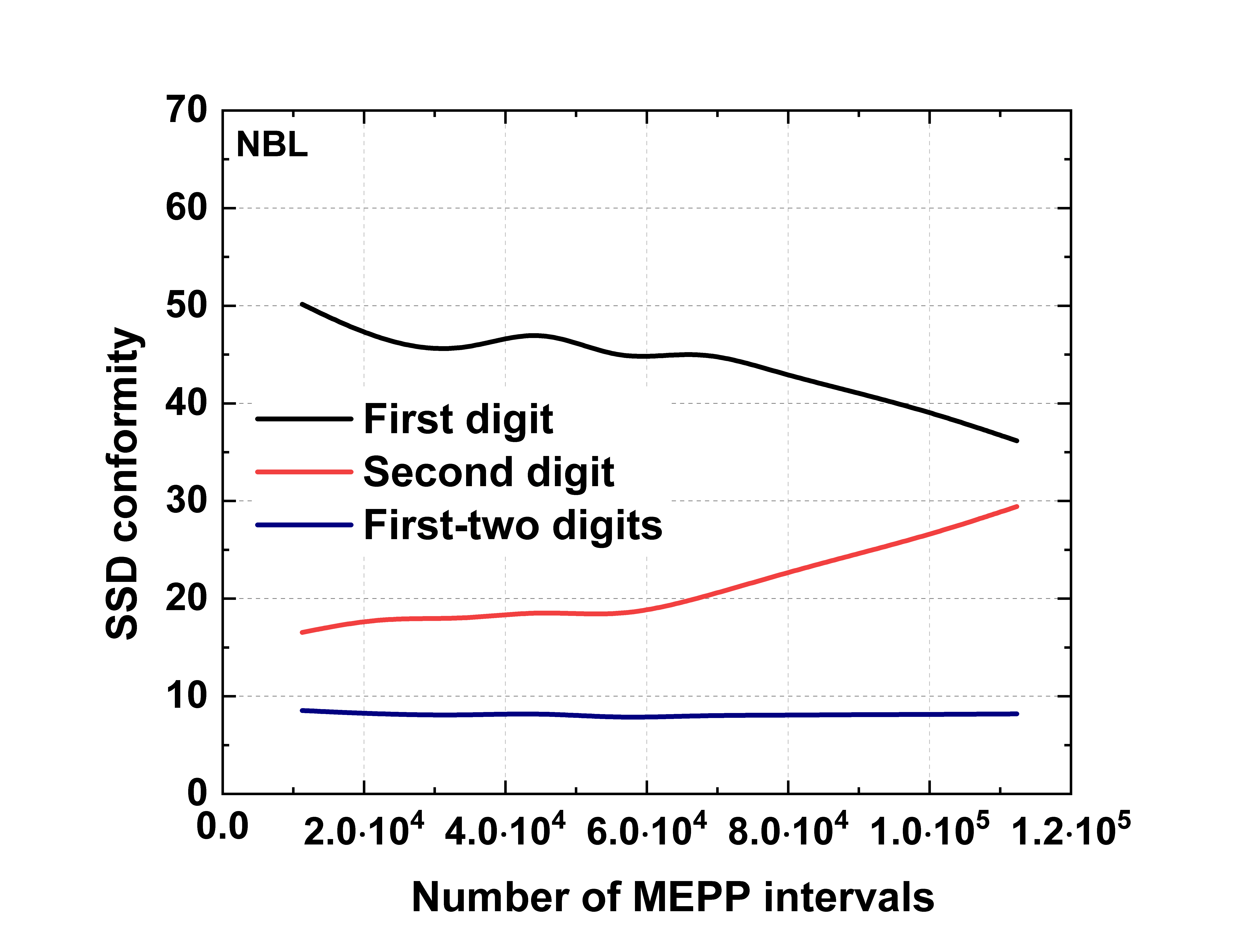}
\includegraphics[width=7.7 cm]{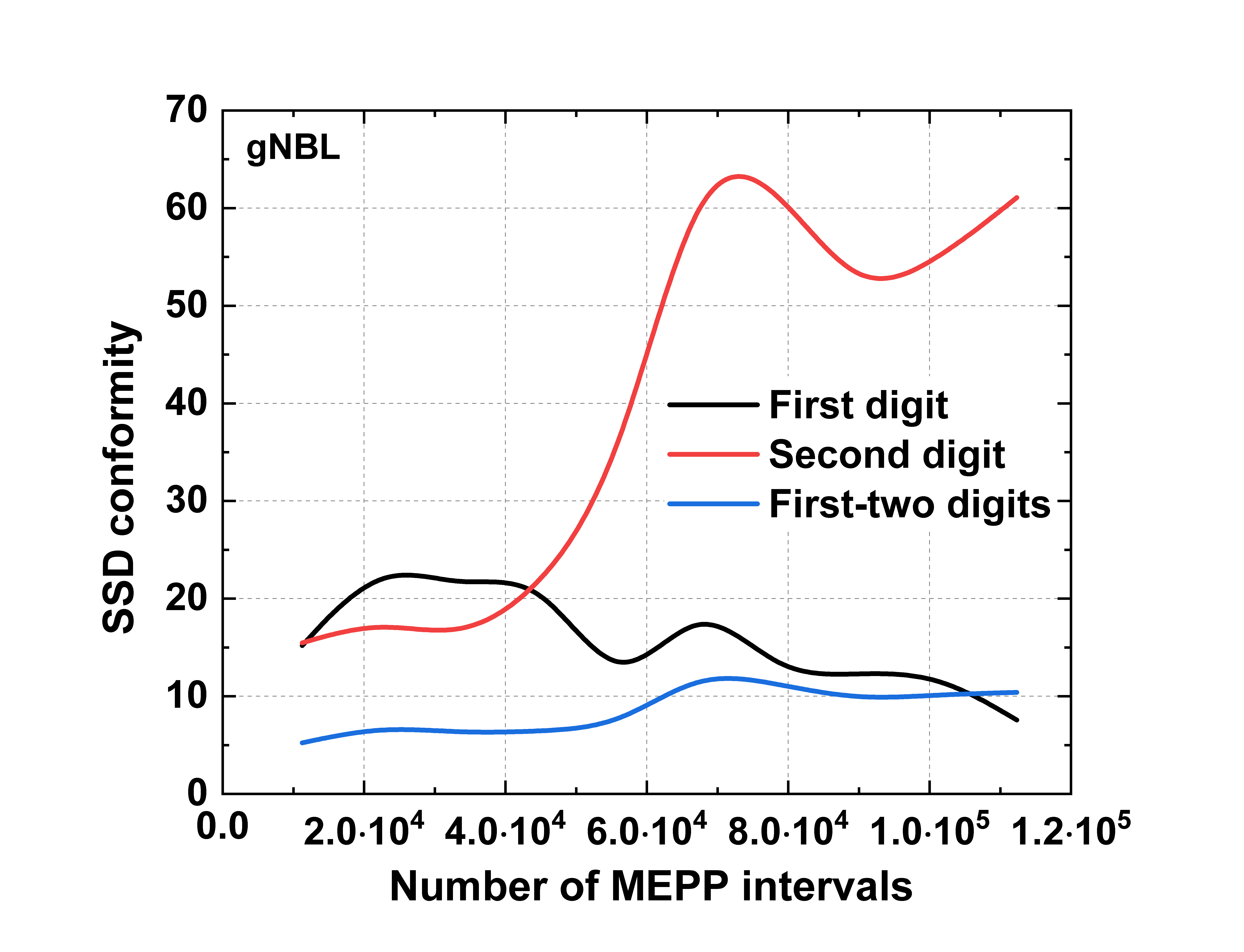}

\includegraphics[width=7.7 cm]{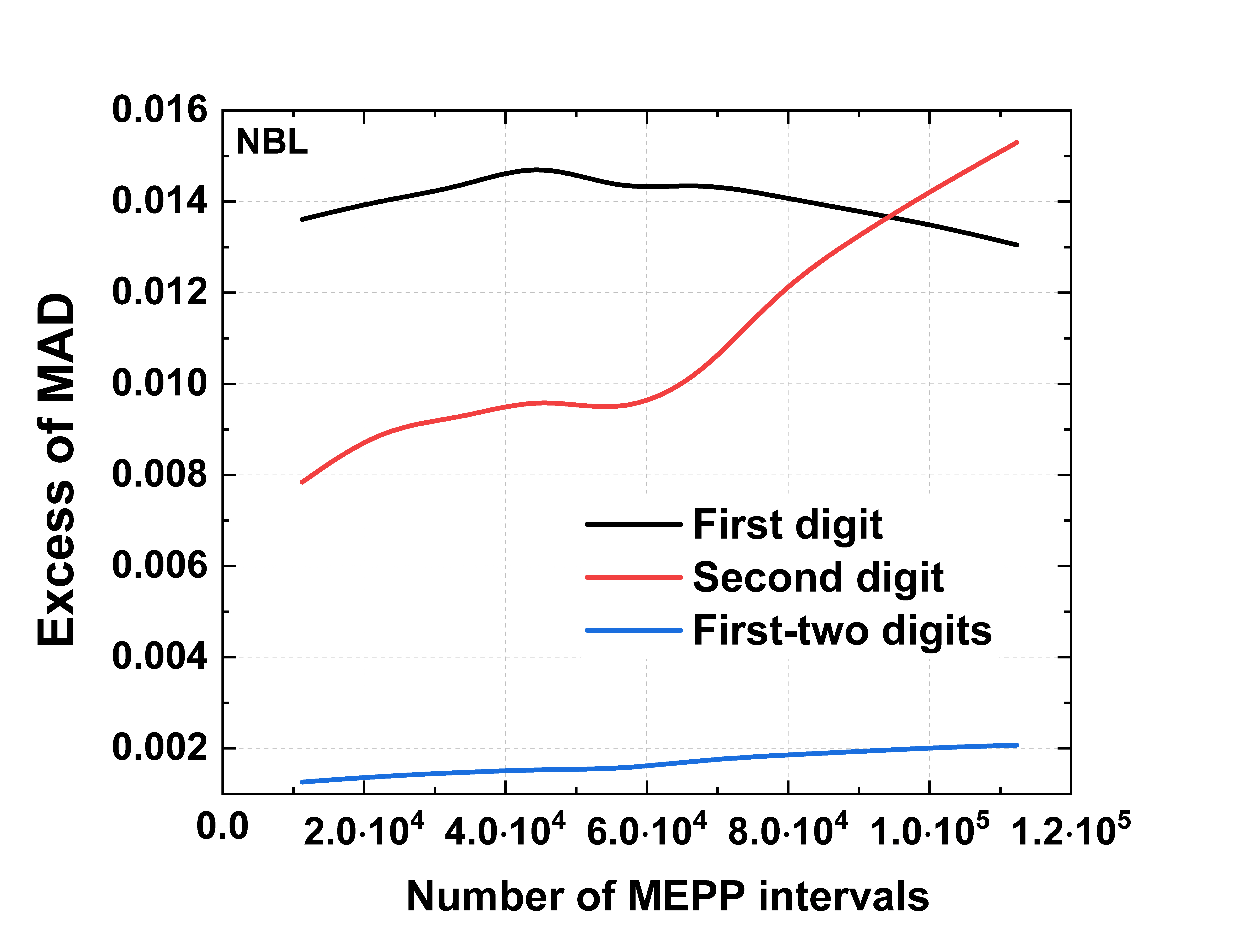}
\includegraphics[width=7.7 cm]{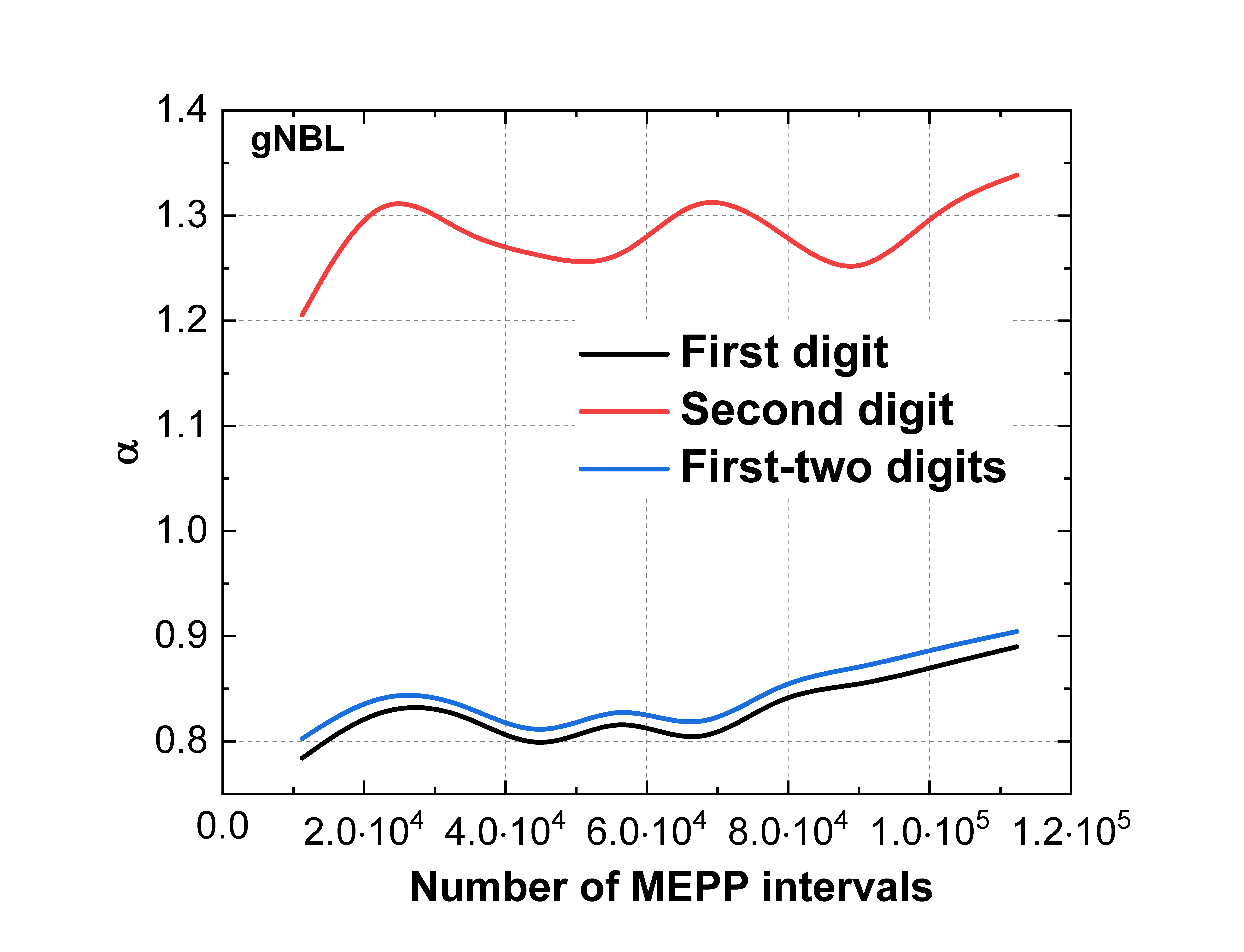}

\caption{Fluctuations in both conformity level and $\alpha$ index.\label{fig4}}
\end{figure}

The high number of intervals obtained in some electrophysiological recordings at high $[K^{+}]_{o}$ offered the possibility of investigating how the level of compliance could be regulated as a function of the number of $MEPP$ intervals. To address this issue, we performed an analysis based on the data cumulative frequency. This procedure divided the experimental data into equal portions within the time series. Next, we successively calculated the compliance level for each data portion. Figure \ref{fig4}, illustrates an analysis made from $112,328$ intervals, which gives the conformity behavior considering the adopted tests. In general, for both $[K^{+}]_{o}$, the tests revealed fluctuations given by the presence of local conformities and non-conformities until achieving the final value giving the final compliance level. It is worth mentioning that this fluctuation was noted in all data, being more evident in those obtained (results not shown).

This calculation enabled us to glimpse how the data amount may contribute to regulating compliance. The data studied with the $NBL$, yielded a gradual improvement in the first digit compliance level. This behavior was analogously observed when $gNBL$ is used, where compliance also improved despite the more pronounced oscillation compared to $NBL$ result, along with the increase in the amount of successively added portions.
 For the second digit, there is a tendency for the level of compliance to increase and to be more unstable of the $MEPP$ intervals quantity. On the other hand, the results for compliance of the first two digits presented a slight variation, suggesting a more attenuated sensitivity concerning the data size.
 Furthermore, our calculations revealed that the $\alpha$ exponent also demonstrated an interesting oscillatory behavior, especially for the second digit compared to the first and two first digits.
 Regarding the tests used to verify the conformity of the experimental data with the $gNBL$, a significant variation in the results of the second digit can be noticed, confirming the tendency observed in the $NBL$ analysis.
 In addition, adopting the $gNBL$ improved the compliance level. In summary, these results strongly suggest that at least for $MEPP$ intervals recorded at mammalian $NMJ$, the conformity level may be modulated by the $MEPP$ frequency or the time series length.

\begin{figure}[!hb]
\centering

\includegraphics[width=9.5 cm]{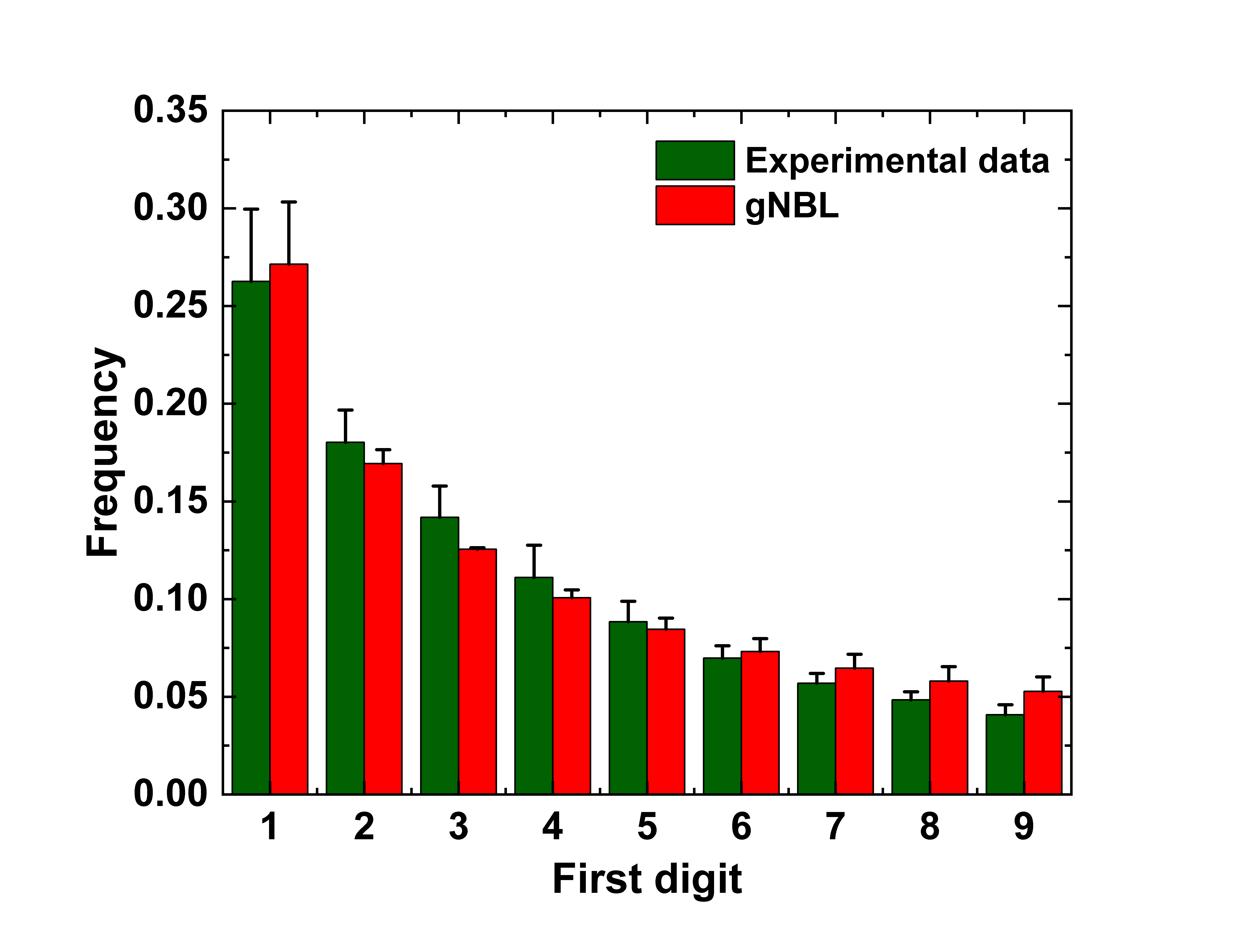}

\includegraphics[width=9.5 cm]{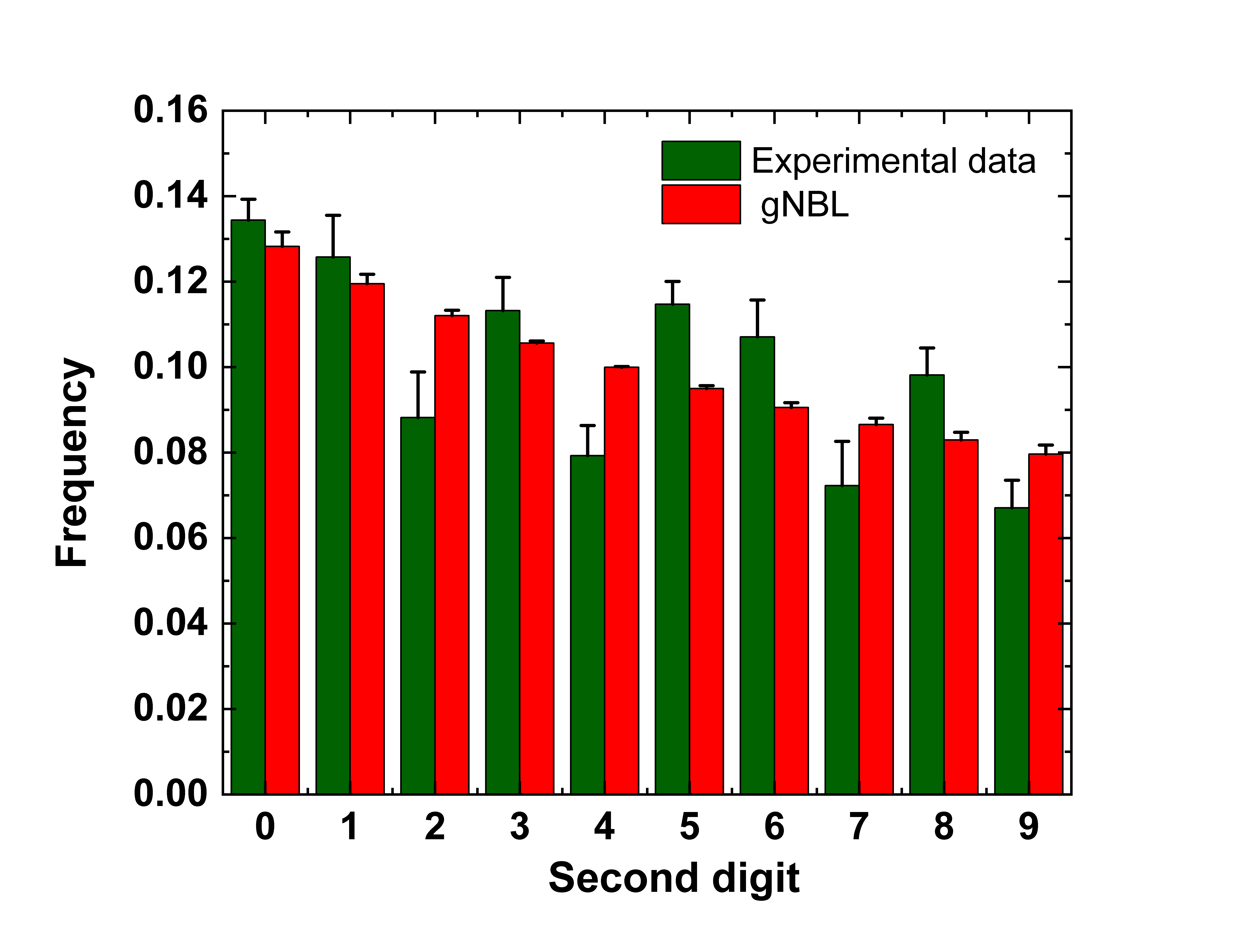}

\includegraphics[width=9.5 cm]{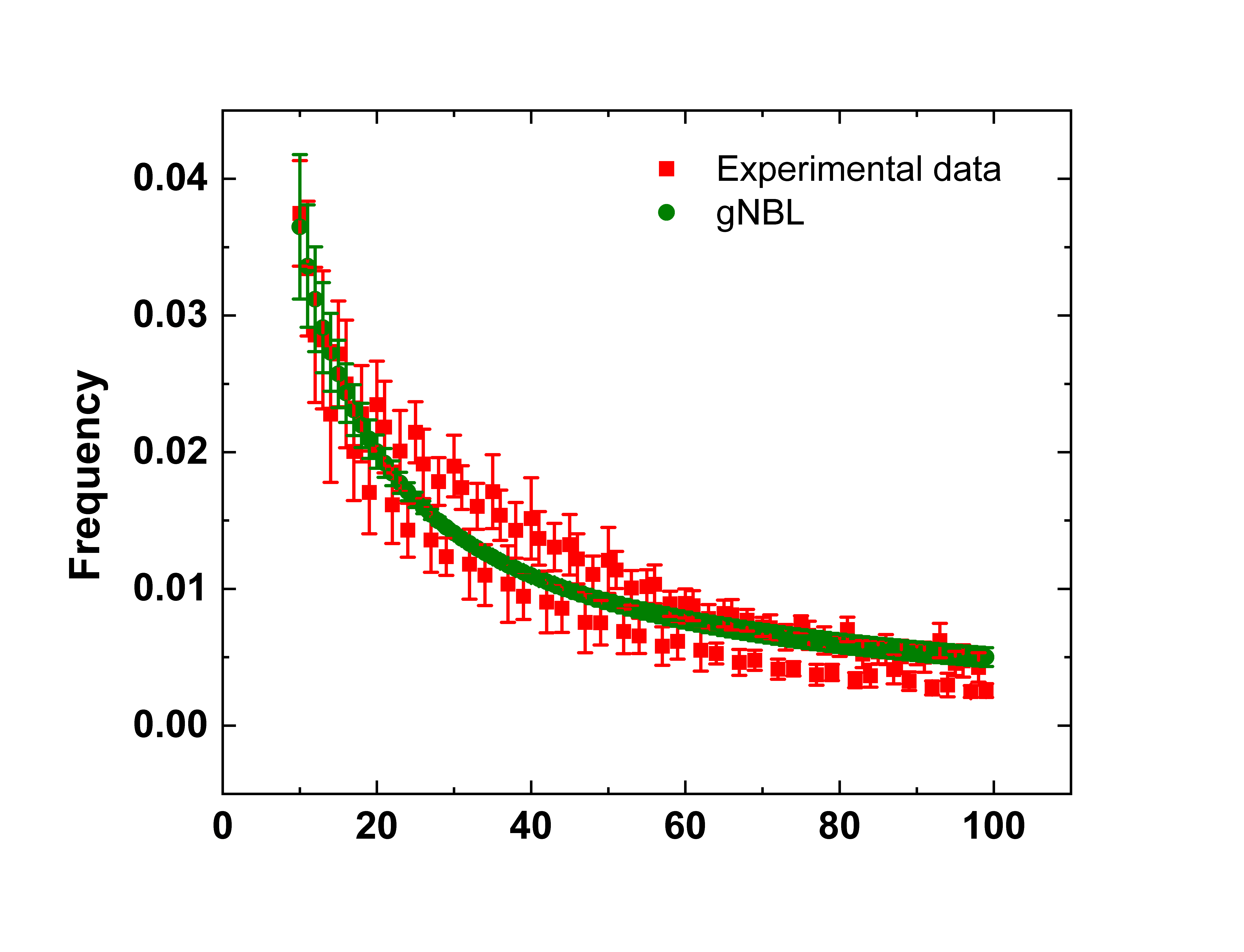}

\caption{Statistical summary for $[K^{+}]_{o} = 25$ mM (n = $12$), using $gNBL$. Values are expressed as mean $\pm$  standard deviation. \label{fig5}}
\end{figure}

\begin{figure}[!hb]
\centering
\hspace{-5.2cm} a)  \hspace{8.cm} b)

\includegraphics[width=7.7 cm]{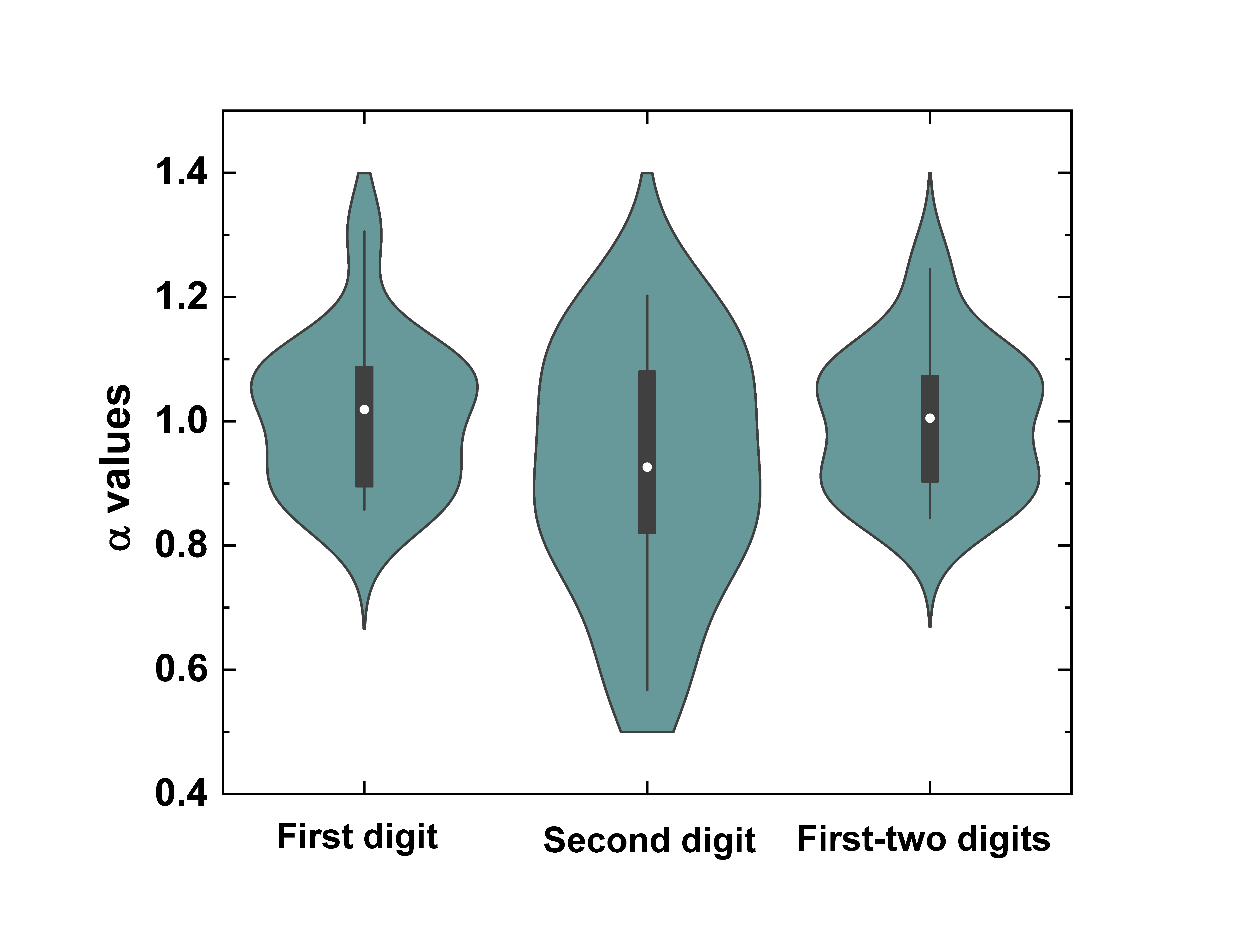}
\includegraphics[width=7.7 cm]{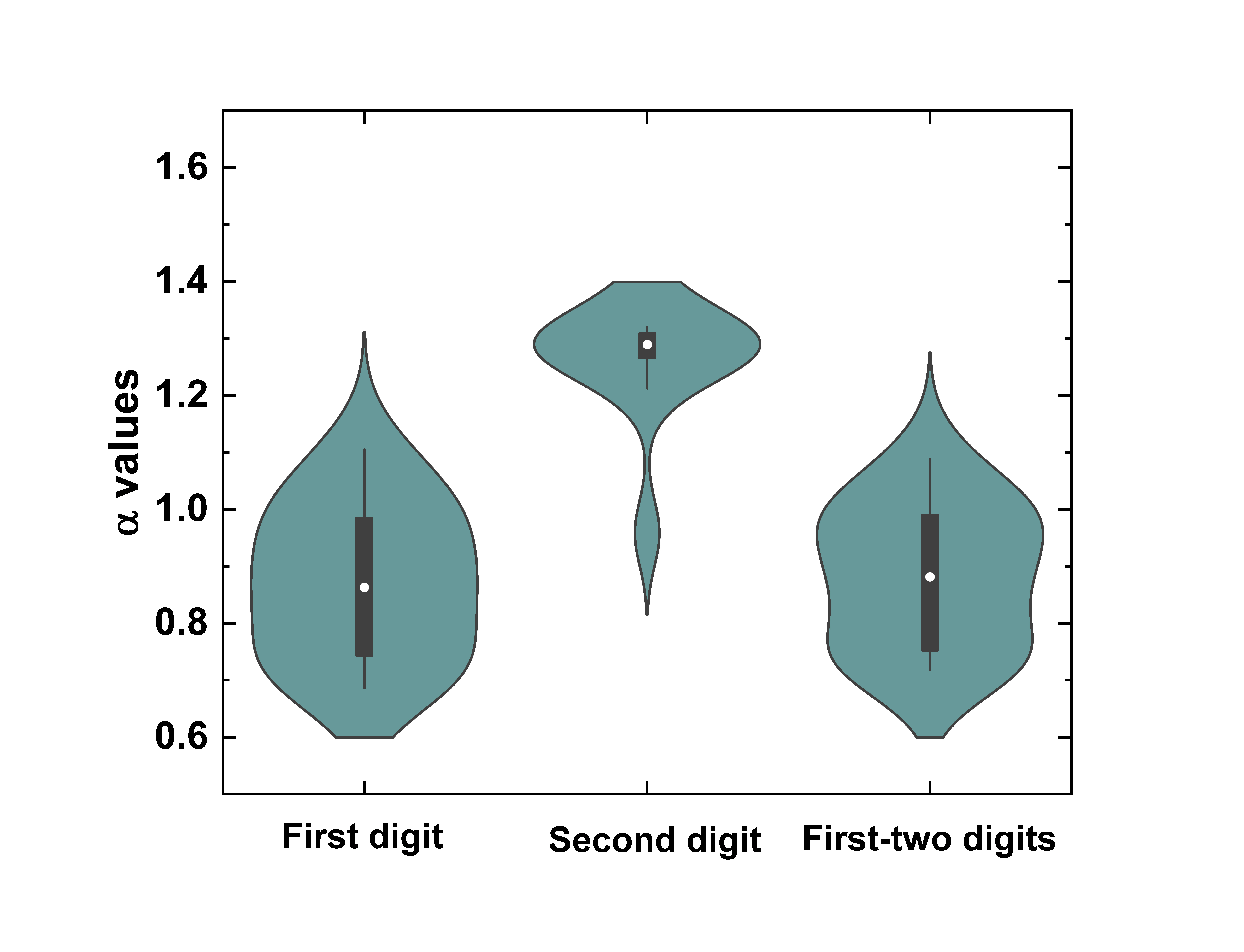}

\caption{Statistical summary for $\alpha$ parameter considering for $[K^{+}]_{o}$ = $5$ mM (n = 14) (a) and $[K^{+}]_{o}$ = $25$ mM (n = 12) (b). \label{fig6}}
\end{figure}

The most pronounced deviations observed at $[K^{+}]_{o}$ in the statistical summary presented in Figure \ref{fig3} motivated us to verify the data disposition considering the $gNBL$. Figure \ref{fig5} provides the statistical summary comparing the distribution of all digits of the experiments carried out in high potassium, in which we can verify a satisfactory enhancement of the $gNBL$ with the experimental data. This visual conformity is seen for the first and first two digits. The $gNBL$ relevance also highlighted by examining the exponent factor $\alpha$, in which values close to $1$, taken at physiological at $[K^{+}]_{o}$, indicates that $NBL$ is sufficiently satisfactory to describe the $MEPP$ intervals (Figure \ref{fig6}). Moreover, based on $\alpha$, one may indicate $gNBL$ importance in modeling data extracted from high $[K^{+}]_{o}$.

\begin{table}[ht]

\caption{Summary of $NBL$ tests for $[K^{+}]_{o}$ = $5$ mM.  \label{tab3}}

\vspace{0.5cm} \small
\hspace{-1.0cm}\begin{tabular}{ccccc}
\toprule
\multicolumn{5}{c}{First Digit}  \\ \midrule
Data	&	n	&	MAD	&	Excess MAD	&	SSD	\\ \midrule

1	&	730	&	0.00837	&	-0.00032	&	12.28475	\\ \midrule
2	&	869	&	0.01995	&	0.01198	&	86.45288	\\ \midrule
3	&	2181	&	0.00989	&	0.00486	&	16.56178	\\ \midrule
4	&	928	&	0.00858	&	0.00087	&	10.08629	\\ \midrule
5	&	2973	&	0.01305	&	0.00874	&	21.54770	\\ \midrule
6	&	642	&	0.00986	&	0.00060	&	9.91123	\\ \midrule
7	&	1349	&	0.00887	&	0.00248	&	10.38188	\\ \midrule
8	&	1162	&	0.01270	&	0.00581	&	18.13111	\\ \midrule
9	&	1685	&	0.00845	&	0.00273	&	11.65409	\\ \midrule
10	&	1009	&	0.01137	&	0.00398	&	16.06490	\\ \midrule
11	&	2048	&	0.00807	&	0.00288	&	7.53752	\\ \midrule
12	&	3060	&	0.00722	&	0.00298	&	8.61009	\\ \midrule
13	&	1510	&	0.01192	&	0.00588	&	27.89886	\\ \midrule
14	&	2436	&	0.00916	&	0.00440	&	9.96616	\\ 

\bottomrule
		\end{tabular} \hspace{1.5cm}
		\begin{tabular}{ccccc}
\toprule
\multicolumn{5}{c}{Second Digit}  \\ \midrule

Data	&	n	&	MAD	&	Excess MAD	&	SSD	\\ \midrule
1	&	730	&	0.00683	&	-0.00202	&	8.19079	\\ \midrule
2	&	869	&	0.00950	&	0.00139	&	12.32210	\\ \midrule
3	&	2181	&	0.00544	&	0.00033	&	4.28136	\\ \midrule
4	&	928	&	0.00851	&	0.00067	&	12.11295	\\ \midrule
5	&	2973	&	0.00318	&	-0.00120	&	1.80299	\\ \midrule
6	&	642	&	0.00558	&	-0.00385	&	4.44022	\\ \midrule
7	&	1349	&	0.00466	&	-0.00184	&	3.15052	\\ \midrule
8	&	1162	&	0.00785	&	0.00084	&	8.96622	\\ \midrule
9	&	1685	&	0.00692	&	0.00110	&	7.73984	\\ \midrule
10	&	1009	&	0.00494	&	-0.00258	&	3.72527	\\ \midrule
11	&	2048	&	0.00525	&	-0.00003	&	3.96622	\\ \midrule
12	&	3060	&	0.00495	&	0.00063	&	3.65075	\\ \midrule
13	&	1510	&	0.00612	&	-0.00003	&	5.18546	\\ \midrule
14	&	2436	&	0.00529	&	0.00045	&	4.44622	\\

\bottomrule
		\end{tabular}

\vspace{0.5cm}  \hspace{5.0cm}	\begin{tabular}{ccccc}
\toprule
\multicolumn{5}{c}{First and Second Digits}  \\ \midrule

Data	&	n	&	MAD	&	Excess MAD	&	SSD	\\ \midrule
1	&	730	&	0.00253	&	-0.00040	&	10.83412	\\ \midrule
2	&	869	&	0.00366	&	0.00097	&	21.72249	\\ \midrule
3	&	2181	&	0.00232	&	0.00062	&	8.81407	\\ \midrule
4	&	928	&	0.00276	&	0.00016	&	12.93240	\\ \midrule
5	&	2973	&	0.00177	&	0.00031	&	5.02102	\\ \midrule
6	&	642	&	0.00268	&	-0.00045	&	9.19816	\\ \midrule
7	&	1349	&	0.00225	&	0.00009	&	7.40826	\\ \midrule
8	&	1162	&	0.00244	&	0.00012	&	8.78088	\\ \midrule
9	&	1685	&	0.00225	&	0.00031	&	8.69891	\\ \midrule
10	&	1009	&	0.00240	&	-0.00010	&	7.84890	\\ \midrule
11	&	2048	&	0.00169	&	-0.00006	&	4.39289	\\ \midrule
12	&	3060	&	0.00141	&	-0.00002	&	4.12983	\\ \midrule
13	&	1510	&	0.00263	&	0.00059	&	10.24864	\\ \midrule
14	&	2436	&	0.00167	&	0.00007	&	4.34060	\\ 
\bottomrule
		\end{tabular}


\end{table}


\begin{table}[ht]
\caption{Summary of $NBL$ tests for $[K^{+}]_{o} = 25$ mM. \label{tab4}}


\vspace{0.5cm} \small
\hspace{-1.8cm} \begin{tabular}{ccccc}
\toprule
\multicolumn{5}{c}{First Digit}  \\ \midrule

Data	&	n	&	MAD	&	Excess MAD	&	SSD	\\ \midrule
1	&	56853	&	0.02687	&	0.02589	&	117.77678	\\ \midrule
2	&	40951	&	0.02383	&	0.02267	&	102.72734	\\ \midrule
3	&	72519	&	0.01827	&	0.01740	&	69.91262	\\ \midrule
4	&	51207	&	0.00910	&	0.00807	&	11.41064	\\ \midrule
5	&	112330	&	0.01375	&	0.01305	&	36.17818	\\ \midrule
6	&	107801	&	0.01952	&	0.01881	&	52.04470	\\ \midrule
7	&	36330	&	0.00619	&	0.00496	&	4.52025	\\ \midrule
8	&	37758	&	0.00117	&	-0.00004	&	0.25865	\\ \midrule
9	&	37703	&	0.00732	&	0.00611	&	7.98955	\\ \midrule
10	&	72787	&	0.02064	&	0.01977	&	92.09016	\\ \midrule
11	&	34833	&	0.01054	&	0.00928	&	19.34788	\\ \midrule
12	&	29313	&	0.00318	&	0.00180	&	1.77512	\\ 

\bottomrule
		\end{tabular} \hspace{1.5cm}
		\begin{tabular}{ccccc}
\toprule
\multicolumn{5}{c}{Second Digit}  \\ \midrule

Data	&	n	&	MAD	&	Excess MAD	&	SSD	\\ \midrule
1	&	56853	&	0.00995	&	0.00895	&	16.15754	\\ \midrule
2	&	40951	&	0.00705	&	0.00586	&	6.48468	\\ \midrule
3	&	72519	&	0.02088	&	0.02000	&	46.27171	\\ \midrule
4	&	51207	&	0.01589	&	0.01484	&	26.70868	\\ \midrule
5	&	112330	&	0.01601	&	0.01530	&	29.43487	\\ \midrule
6	&	107801	&	0.02213	&	0.02140	&	52.58192	\\ \midrule
7	&	36330	&	0.01576	&	0.01450	&	26.28400	\\ \midrule
8	&	37758	&	0.01441	&	0.01318	&	21.98806	\\ \midrule
9	&	37703	&	0.01253	&	0.01130	&	16.74141	\\ \midrule
10	&	72787	&	0.02409	&	0.02320	&	61.55550	\\ \midrule
11	&	34833	&	0.01854	&	0.01726	&	37.01945	\\ \midrule
12	&	29313	&	0.01897	&	0.01757	&	38.09396	\\ 

\bottomrule
		\end{tabular}

\vspace{1.0cm}  \hspace{5.0cm}	\begin{tabular}{ccccc}
\toprule
\multicolumn{5}{c}{First and Second Digits}  \\ \midrule

Data	&	n	&	MAD	&	Excess MAD	&	SSD	\\ \midrule
1	&	56853	&	0.00301	&	0.00268	&	15.60049	\\ \midrule
2	&	40951	&	0.00249	&	0.00209	&	12.89166	\\ \midrule
3	&	72519	&	0.00308	&	0.00278	&	13.89474	\\ \midrule
4	&	51207	&	0.00181	&	0.00146	&	4.98217	\\ \midrule
5	&	112330	&	0.00231	&	0.00207	&	8.19403	\\ \midrule
6	&	107801	&	0.00318	&	0.00294	&	14.78929	\\ \midrule
7	&	36330	&	0.00192	&	0.00150	&	3.94064	\\ \midrule
8	&	37758	&	0.00163	&	0.00122	&	2.85762	\\ \midrule
9	&	37703	&	0.00160	&	0.00119	&	3.65519	\\ \midrule
10	&	72787	&	0.00349	&	0.00320	&	17.89426	\\ \midrule
11	&	34833	&	0.00243	&	0.00200	&	6.71510	\\ \midrule
12	&	29313	&	0.00217	&	0.00171	&	5.13421	\\ 

\bottomrule
		\end{tabular}


\end{table}


\begin{table}[ht]
\caption{Summary of $gNBL$ tests for $[K^{+}]_{o}$ = $5$ mM. \label{tab5}}

\vspace{0.5cm} \small
\hspace{-0.4cm} \begin{tabular}{cccccc}
\toprule
\multicolumn{5}{c}{First Digit}  \\ \midrule

Data	&	n	&	Alpha	&	MAD	&	SSD	\\ \midrule
1	&	730	&	1,09734	&	0,00636	&	5,35810	\\ \midrule
2	&	869	&	1,30536	&	0,01065	&	16,90700	\\ \midrule
3	&	2181	&	0,85783	&	0,00349	&	1,97250	\\ \midrule
4	&	928	&	0,98698	&	0,00828	&	9,96370	\\ \midrule
5	&	2973	&	0,87338	&	0,00851	&	9,87620	\\ \midrule
6	&	642	&	1,07180	&	0,00713	&	5,96820	\\ \midrule
7	&	1349	&	0,96904	&	0,00844	&	9,65490	\\ \midrule
8	&	1162	&	1,08773	&	0,01008	&	12,10020	\\ \midrule
9	&	1685	&	0,93251	&	0,00728	&	8,23990	\\ \midrule
10	&	1009	&	0,89552	&	0,00824	&	8,04460	\\ \midrule
11	&	2048	&	1,05146	&	0,00661	&	5,51620	\\ \midrule
12	&	3060	&	1,09466	&	0,00368	&	1,97250	\\ \midrule
13	&	1510	&	0,86705	&	0,01149	&	14,54340	\\ \midrule
14	&	2436	&	1,06165	&	0,00794	&	7,02810	\\ 

\bottomrule
		\end{tabular} \hspace{1.5cm}
		\begin{tabular}{cccccc}
\toprule
\multicolumn{5}{c}{Second Digit}  \\ \midrule

Data	&	n	&	Alpha	&	MAD	&	SSD	\\ \midrule
1	&	730	&	0,88315	&	0,00703	&	7,88420	\\ \midrule
2	&	869	&	1,20207	&	0,00977	&	11,27090	\\ \midrule
3	&	2181	&	0,76299	&	0,00480	&	3,19950	\\ \midrule
4	&	928	&	0,56741	&	0,00699	&	8,74690	\\ \midrule
5	&	2973	&	0,96940	&	0,00328	&	1,78210	\\ \midrule
6	&	642	&	0,82057	&	0,00565	&	3,79340	\\ \midrule
7	&	1349	&	0,83357	&	0,00431	&	2,57830	\\ \midrule
8	&	1162	&	1,14257	&	0,00794	&	8,44120	\\ \midrule
9	&	1685	&	1,14975	&	0,00653	&	7,19650	\\ \midrule
10	&	1009	&	1,04217	&	0,00485	&	3,68420	\\ \midrule
11	&	2048	&	0,82621	&	0,00465	&	3,33970	\\ \midrule
12	&	3060	&	1,08013	&	0,00531	&	3,49580	\\ \midrule
13	&	1510	&	0,61821	&	0,00425	&	2,57810	\\ \midrule
14	&	2436	&	0,98519	&	0,00538	&	4,44090	\\ 

\bottomrule
		\end{tabular}

\vspace{1.0cm}  \hspace{5.0cm}	\begin{tabular}{cccccc}
\toprule
\multicolumn{5}{c}{First and Second Digits}  \\ \midrule

Data	&	n	&	Alpha	&	MAD	&	SSD	\\ \midrule
1	&	730	&	1,11785	&	0,00241	&	9,63330	\\ \midrule
2	&	869	&	1,24452	&	0,00329	&	16,11180	\\ \midrule
3	&	2181	&	0,86327	&	0,00214	&	7,27660	\\ \midrule
4	&	928	&	0,97734	&	0,00276	&	12,88900	\\ \midrule
5	&	2973	&	0,87317	&	0,00157	&	3,66930	\\ \midrule
6	&	642	&	1,06433	&	0,00259	&	8,82400	\\ \midrule
7	&	1349	&	0,94217	&	0,00219	&	7,11010	\\ \midrule
8	&	1162	&	1,07238	&	0,00236	&	8,29090	\\ \midrule
9	&	1685	&	0,91808	&	0,00223	&	8,10760	\\ \midrule
10	&	1009	&	0,90327	&	0,00230	&	7,05690	\\ \midrule
11	&	2048	&	1,05125	&	0,00166	&	4,15790	\\ \midrule
12	&	3060	&	1,09810	&	0,00129	&	3,28490	\\ \midrule
13	&	1510	&	0,84443	&	0,00248	&	8,13420	\\ \midrule
14	&	2436	&	1,03307	&	0,00165	&	4,23940	\\ 

\bottomrule
		\end{tabular}


\end{table}

\begin{table}[ht]
\caption{Summary of $gNBL$ tests for $[K^{+}]_{o} = 25$ mM.  \label{tab6}}


\vspace{0.5cm} \small
\hspace{-0.8cm} 		\begin{tabular}{cccccc}
\toprule
\multicolumn{5}{c}{First Digit}  \\ \midrule

Data	&	n	&	Alpha	&	MAD	&	SSD	\\ \midrule
1	&	56853	&	0,68616	&	0,01982	&	45,09730	\\ \midrule
2	&	40951	&	0,73411	&	0,02093	&	47,66000	\\ \midrule
3	&	72519	&	0,75305	&	0,01487	&	23,73950	\\ \midrule
4	&	51207	&	1,10501	&	0,00427	&	3,05230	\\ \midrule
5	&	112330	&	0,83126	&	0,01152	&	14,63950	\\ \midrule
6	&	107801	&	0,87603	&	0,01860	&	39,84270	\\ \midrule
7	&	36330	&	0,92700	&	0,00228	&	0,68170	\\ \midrule
8	&	37758	&	1,00437	&	0,00123	&	0,24450	\\ \midrule
9	&	37703	&	1,02816	&	0,00712	&	7,37960	\\ \midrule
10	&	72787	&	0,70165	&	0,01540	&	25,51630	\\ \midrule
11	&	34833	&	0,84972	&	0,00475	&	2,89430	\\ \midrule
12	&	29313	&	0,96620	&	0,00265	&	0,92750	\\ 

\bottomrule
		\end{tabular} \hspace{1.5cm}
		\begin{tabular}{cccccc}
\toprule
\multicolumn{5}{c}{Second Digit}  \\ \midrule

Data	&	n	&	Alpha	&	MAD	&	SSD	\\ \midrule
1	&	56853	&	1,26913	&	0,00928	&	14,27730	\\ \midrule
2	&	40951	&	0,95827	&	0,00709	&	6,44590	\\ \midrule
3	&	72519	&	1,26304	&	0,01965	&	44,44780	\\ \midrule
4	&	51207	&	1,30622	&	0,01444	&	24,16300	\\ \midrule
5	&	112330	&	1,28374	&	0,01468	&	27,29740	\\ \midrule
6	&	107801	&	1,32020	&	0,02061	&	49,80040	\\ \midrule
7	&	36330	&	1,29637	&	0,01436	&	23,90470	\\ \midrule
8	&	37758	&	1,27543	&	0,01312	&	19,94300	\\ \midrule
9	&	37703	&	1,21247	&	0,01155	&	15,56480	\\ \midrule
10	&	72787	&	1,29532	&	0,02269	&	59,22550	\\ \midrule
11	&	34833	&	1,38829	&	0,01666	&	32,80310	\\ \midrule
12	&	29313	&	1,31114	&	0,01749	&	35,45490	\\ 

\bottomrule
		\end{tabular}

\vspace{1.0cm}  \hspace{5.0cm}	\begin{tabular}{cccccc}
\toprule
\multicolumn{5}{c}{First and Second Digits}  \\ \midrule

Data	&	n	&	Alpha	&	MAD	&	SSD	\\ \midrule
1	&	56853	&	0,71868	&	0,00249	&	9,14370	\\ \midrule
2	&	40951	&	0,73705	&	0,00220	&	6,82500	\\ \midrule
3	&	72519	&	0,76779	&	0,00272	&	9,27860	\\ \midrule
4	&	51207	&	1,08779	&	0,00183	&	4,27770	\\ \midrule
5	&	112330	&	0,84541	&	0,00215	&	6,12730	\\ \midrule
6	&	107801	&	0,87479	&	0,00314	&	13,35130	\\ \midrule
7	&	36330	&	0,95909	&	0,00189	&	3,80170	\\ \midrule
8	&	37758	&	1,00499	&	0,00164	&	2,85540	\\ \midrule
9	&	37703	&	1,00683	&	0,00160	&	3,65090	\\ \midrule
10	&	72787	&	0,72606	&	0,00309	&	11,62400	\\ \midrule
11	&	34833	&	0,88760	&	0,00224	&	5,66890	\\ \midrule
12	&	29313	&	0,97453	&	0,00214	&	5,07780	\\ 

\bottomrule
		\end{tabular}


\end{table}

\section{Discussion}

The present study expanded our previous investigation on how changes in the ionic concentration of the artificial physiological solution can modulate the level of compliance of intervals between $MEPPs$. In this framework, this report confirmed the $NBL$ validity in a hyperkalemic environment. As already expected, the analysis initially showed that intervals of $MEPP$, recorded at normal $[K^{+}]_{o}$, agree with the first, second, and first two digits frequencies. We achieved this conclusion by assuming three different conformity tests. In $[K^{+}]_{o}$ = 5 mM, excess $MAD$ test enabled improved conformity for both first and second digit results. At the same time, for the first two digits data, all nonconformities were converted into conformance. These findings suggest how "revealed as "excess power" may influence the results and data interpretation.

According to our analysis, at the $[K^{+}]_{o}$ = 25 mM, a heterogeneous conformity scenario emerged, in which nonconformity abundantly appeared as compared to the results at physiological solution. Besides this observation, $SSD$ generally pointed out that the distribution of digits obeys the $NBL$. The level of compliance predicted by this test in most cases examined relied on an acceptable and marginal level. On the other hand, the results obtained from $MAD$ and excess $MAD$ calculations provided several nonconformities, showing the necessity to adopt a generalized $NBL$ version to understand applicability and limitations for $[K^{+}]_{o}$ = 25 mM. Also, the strong depolarization, promoted by the high $[K^{+}]_{o}$, resulted in the loss of conformity. In this framework, $gNBL$ assumption brings closer the $MEPP$ intervals adherence to the law, especially when quantified by the $SSD$ test. In addition, $\alpha$ values emerged as a helpful parameter for verifying if the data is better described by the $gNBL$ or $NBL$. When $gNBL$ is assumed, the results for $[K^{+}]_{o}$ within the physiological content showed that $\alpha$ median is close to $1$. In contrast, at high $[K^{+}]_{o}$, they deviate more significantly from $\alpha = 1$, highlighting the importance of $gNBL$ in data harvested at higher concentrations.

Based on the present findings, one may formulate the following question: What neural substrate might be associated with law validation at high $[K^{+}]_{o}$? Is there a relationship between morphological modifications and the rate of discharge of $MEPPs$? Therefore, is the $gNBL$, given by its $\alpha$ parameter, a possible indicator of the structural changes in the $NMJ$? Previous research suggested that high $[K^{+}]_{o}$ is related to morphological alterations as much as observed in a series of pathologies. Although our approach to tackle this question was indirect, focusing only on the electrical response, further combinations of morphological and electrophysiological studies are required to investigate how changes in $NMJ$ morphology can be associated with $\alpha$ exponent. Yet, within this scope, it would also be essential to investigate the validity of the law in situations of injury, in which it is well accepted that the $NMJ$ terminal undergoes morphological restructuring as well. Examination of these issues may confirm the utility of $gNBL$ in quantifying numerical patterns in pathologies known for modifying the $NMJ$ architecture. If such correspondence could be finally confirmed, $gNBL$ would arise as a suitable form for detecting the presence of an anomalous regime beyond those associated with hyperkalemia diseases.

In this work, the experiments were done considering the ambient temperature. Besides $[K^{+}]_{o}$ increment, the thermal level arising is another critical parameter that promotes $MEPP$ frequency modulations. It is evident that thermal fluctuations modify the resting potential of the nerve terminal, depolarizing and hyperpolarizing as the biological membrane temperature is raised and lowered, respectively \cite{nakanishi1969}. Consequently, although changes in temperature and $[K^{+}]_{o}$ imply different synaptic mechanisms, rising temperature similarly reflects an increment of $MEPP$ frequency \cite{ward1972,white1976}. Thus, one may hypothesize that at higher temperatures, such as observed for a hyperkalemic environment, $gNBL$ would emerge as the most appropriate formulation to study the first digit phenomenon at mammalian physiological temperature. In that case, it is plausible to argue that resting potential might be governed by physiological mechanisms ruled by generalized formalism. This conjecture is based on the following arguments. Firstly, Procópio and Fornés, inspired by the fluctuation-dissipation theorem, showed how voltage fluctuations impose a mechanism responsible for regulating the gating channel behavior \cite{procopio}. Secondly, influenced by generalized thermodynamics statistics ($GTS$), Chame and Mello generalized the fluctuation-dissipation theorem \cite{chame}. Third, a direct mathematical relation between the $NBL$ and $GTS$ was deduced by Shao and Ma \cite{shao}. Finally, studies at mammalian $NMJ$ performed by da Silva \textit{et al.} have shown that synaptic transmission statistics are best understood within an approach inspired by the $GTS$ \cite{silva2015,floquet}. Taken together, these arguments sketch a theoretical pillar to hypothesize the existence of a relation between $gNBL$ within a generalized resting potential, likely valid at mammalian temperatures. In this scheme, $\alpha \neq 1$ would imply a resting potential regulated in terms of $GTS$ formalism and its famous $q$-index. Therefore, the discussion given above offers a thermodynamic scenario for explaining the decrement or even conformity fails, computed for the first digits at high $[K^{+}]_{o}$. But, future investigations are required to better comprehend the relationship between the temperature, $NBL$, and $GTS$ theory in the neurotransmission context.

Finally, it is essential to mention that large amounts of data, like those extracted at high $[K^{+}]_{o}$, represent an excellent way to assess how compliance can be changed as a function of the size of the data amount. This especially became more significant for the second and first two digits. Such discussion allows us to elaborate on a last profound question: Will the validity of the law reported here in \textit{in vitro} conditions still be verified at the systemic level, where the junction is intact and attached to the animal organism? Although taken at an artificial hyperkalemic physiological solution, our results showed local variations in the conformity level. Is this compliance behavior sensitive to the sampling rate or the size of the time series? Our results show that, at least at the $NMJ$, the conformity level has a very dynamic behavior. Last, but not least, does the $NBL$ validity change throughout the rodent life? Could these results be equally extrapolated for the human $NMJ$? These are conundrums within the applicability and validity of $NBL$ in physiological terms. Further experimental investigations are welcome to assess these intriguing questions.

\section{Conclusions}

In this work, we examined the $MEEP$ time series to confirm the validity of $NBL$ in hyperkalemic conditions. This strategy validated $NBL$ outside a physiological scenario characterized by an accelerated release of neurotransmitters, artificially induced by manipulating the artificial physiological solution. The assumption of a high potassium concentration allowed us to assess the robustness of the law for a large amount of data. In addition, these results reinforce our previous findings, pointing to a possible ubiquity of the law at the $NMJ$, at least in electrophysiological signals collected from \textit{in vitro} experiments. In the future, we hope to use the same experimental protocols to investigate the validity of this law in electrical signals collected from synapses in the brain. Finally, it is tempting to suggest that, given that the $NBL$ is generally obeyed in different extracellular concentrations of potassium, non-compliance with the $NBL$ could configure its utility in a pathological condition.

\ack

One of the authors (A. J. da Silva) would like to dedicate this article to the memory of Mrs. Maria Pinheiro da Silva and Mrs. Jolinda Pinheiro da Silva for their support and encouragement in his academic career. We are also indebted to the valuable support and suggestions of Alex Ely Kossovsky and prof. Claudio Cerqueti.

\end{document}